\documentclass[11pt]{article}

\usepackage[utf8]{inputenc}
\usepackage[T1]{fontenc}
\usepackage{lmodern}

\usepackage[margin=1in]{geometry}
\usepackage{setspace}
\usepackage{amsmath,amssymb}

\usepackage{booktabs}
\usepackage{array}
\usepackage{tabularx}
\usepackage{longtable}
\usepackage{multirow}

\usepackage{graphicx}
\graphicspath{{photo/}}
\usepackage{float}
\usepackage[export]{adjustbox}
\adjustboxset{max width=0.8\textwidth, max height=0.7\textheight}

\usepackage{titlesec}
\usepackage{enumitem}
\titleformat{\section}{\normalfont\Large\bfseries}{\thesection}{0.6em}{}

\usepackage{xcolor}
\usepackage{fancyvrb}
\usepackage{tikz}
\usetikzlibrary{positioning}

\usepackage[authoryear]{natbib}

\usepackage{caption}
\usepackage[hypertexnames=false]{hyperref}
\hypersetup{
	colorlinks=true,
	linkcolor=blue,
	citecolor=blue,
	urlcolor=blue
}

\title{Are Economists Open to AI? A Text-as-Data-as-Survey Approach via Language Models\thanks{{\footnotesize We appreciate the comments of Iv{a}n Blanco S{a}nchez, Pedro Alberto Chauffaille Saffi, Carlo Chiarella, John Coah, Eric Duca, Pablo Garc{i}a Est{e}vez, Sergio Javier Garc{i}a Saiz, Salom{o}n Garc{i}a Villegas, Juan Pedro Gomez, Manuel L{o}pez Mart{i}n de Blas, Xiaojun Song, Zhijie Xiao, Qingsong Yao, Yonghui Zhang, seminar participants at CUNEF University, University of Bath and  Renmin University of China for generous conversations, careful suggestions, and encouragement at different stages of this project. Their comments helped sharpen the paper's framing and exposition. All remaining errors are our own.}}}

\author{
    Yi Wang\thanks{Renmin University of China. Email: \href{mailto:wy\_139@ruc.edu.cn}{wy\_139@ruc.edu.cn}}
    \and
    Lei Ge\thanks{Renmin University of China. Email: \href{mailto:gelei@ruc.edu.cn}{gelei@ruc.edu.cn}}
}

\date{\today}

\newcommand{\casepair}[2]{\begin{minipage}[t]{\linewidth}\vspace{0pt}\begin{itemize}[leftmargin=1.2em,noitemsep,topsep=0pt]\item #1\item #2\end{itemize}\end{minipage}}

\begin{document}

\maketitle

\begin{abstract}

\noindent Traditional surveys yield comparable measures but are costly to field, difficult to reconstruct retrospectively, and often ill suited to fast-moving or sensitive topics. While large-scale internet text is often noisy and weakly structured. To bridge this gap, we introduce Text-as-Data-as-Survey (TaDaS). TaDaS employs Reference-Anchored Semantic Reparameterization (RAS) to project unstructured main text into survey-like evidence, leveraging structured auxiliary text as semantic anchors. Applying TaDaS to 1.25 million Economics Job Market Rumors posts linked with 53,585 top economics and finance publications, we track economists' evolving research sentiment toward AI. Cross-sectionally, AI-related research discussions are less open, with openness and curiosity declining rapidly at first years. Over time, however, economists have become increasingly open and curious, with a notable shift around 2018. Ultimately, TaDaS provides a scalable, non-reactive method to extract longitudinal insights from digital archives, unlocking diverse applications across industry and academia.
\end{abstract}

\vspace{1em}



\clearpage

\section{Introduction}

Archival text records attitudes, evaluations, and reactions at a scale that surveys rarely match, but it does not supply a shared questionnaire or response scale. Recovering comparable measures from reviews, investor discussions, employee feedback, public comments, and educational text therefore requires more than a generic classifier: it requires a measurement design that states what the semantic dimensions are and why they are comparable across observations \citep{grimmer2013,gentzkow2019}. Surveys remain valuable precisely because their common instrument supplies this structure, although they can be costly, infrequent, retrospective only with difficulty, and sensitive to self-presentation \citep{groves2009,bertrand2001,tourangeau2007}.

TaDaS (Text as Data as Survey) supplies that design for naturally occurring text. Its Reference-Anchored Semantic Reparameterization (RAS) design assigns distinct roles to a structured anchor corpus and an unstructured response corpus. RAS estimates topical semantic anchors from the anchor corpus, then projects response-text embeddings onto those anchors to form low-dimensional topicality coordinates. The intuition is survey-like: a stable question frame is replaced by an explicit, auditable semantic frame, while the archival text supplies the responses. Outcome scoring, human coding, or observed behavior can be added downstream when an application calls for them.

We use EJMR--Artificial Intelligence (AI) as a demanding illustration, not as TaDaS's sole domain. The analysis pairs 53,585 economics and finance publications, which anchor a research-topic space, with 1.25 million posts from Economics Job Market Rumors (EJMR), a noisy archive of naturally occurring discourse. The setting tests whether the framework can preserve interpretable coordinates when the response corpus is anonymous, heterogeneous, and only indirectly related to the anchor corpus.

The paper makes three contributions. First, it offers a measurement architecture for converting naturally occurring text into reusable, survey-like coordinates. Second, it makes semantic anchoring transparent by separating the anchor corpus, anchor construction, response corpus, and any downstream outcome layer. Third, it demonstrates the framework in noisy naturally occurring discourse, showing how the design can organize topicality and associated attitudes without mistaking archival text for a sampled survey. Section \ref{sec:tadas} presents the framework; the remaining sections situate it in the literature, document the EJMR--AI illustration, and report its descriptive results.

\section{Text as Data as Survey}
\label{sec:tadas}

As a benchmark, a conventional survey structure $\mathcal{D}^{survey}$ can be represented as a collection
\begin{equation}
\mathcal{D}^{survey}=\{(y_{i\ell},x_{i\ell},Z_i):i\in\mathcal{I},\ell=1,\dots,k\},
\end{equation}
where $y_{i\ell}$ is the target variable we're interested in, $x_{i\ell}$ is the answer provided by respondent $i$ to the specific questions, and $Z_i$ records respondent or context covariates. We will use $\mathcal{D}^{survey}$ and similar expressions as the notation of survey or survey-like variable structures.

\begin{figure}[H]
    \centering
    \includegraphics[width=\linewidth]{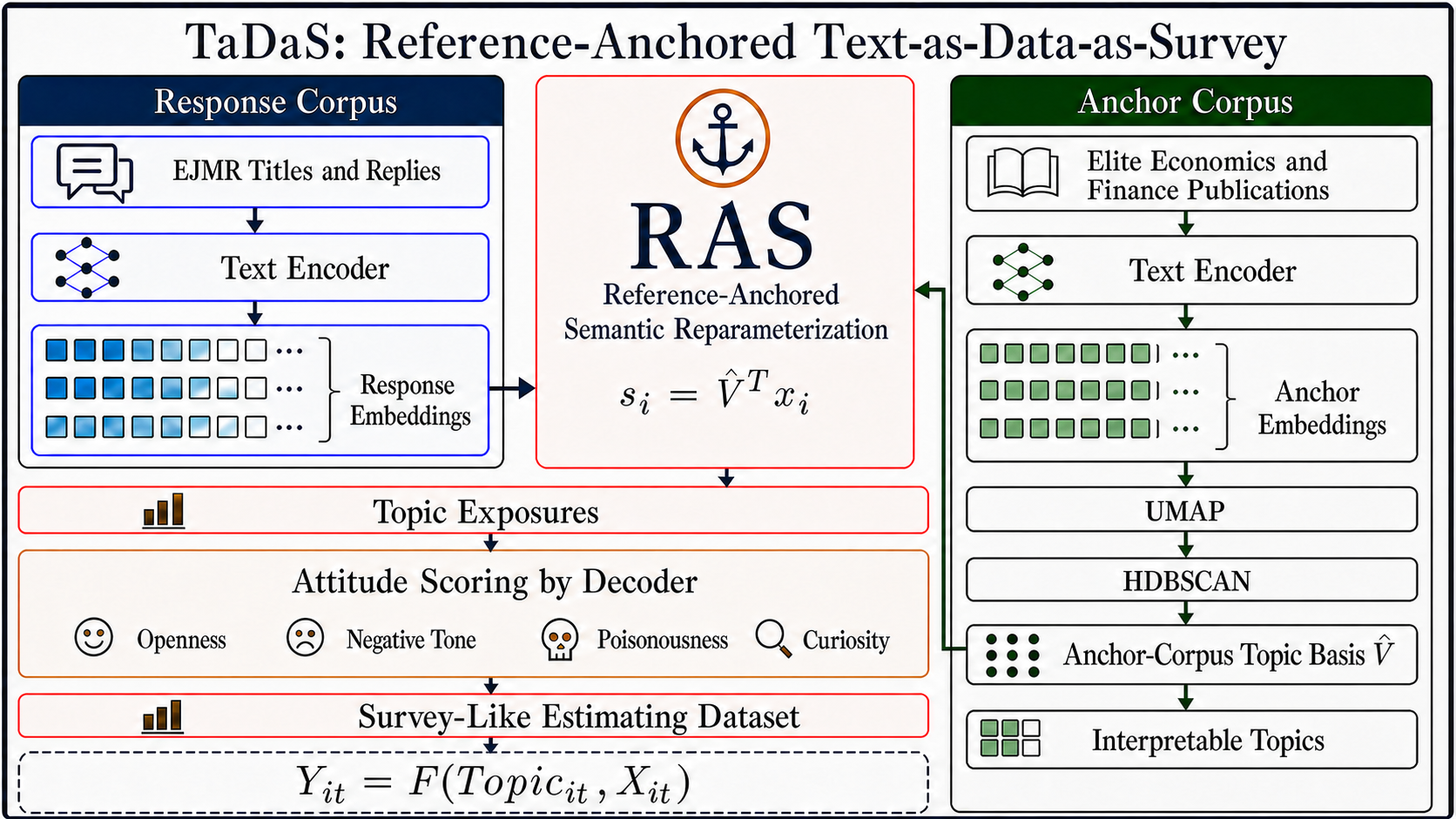}
    \caption{TaDaS: From Anchor Corpora to Survey-Like Measures.}
    \label{fig:workflow}
\end{figure}

TaDaS seeks to construct analogous low-dimensional variables from naturally occurring text. The anchor corpus defines the semantic directions, and the response corpus supplies the text to be measured. Reference-Anchored Semantic Reparameterization (RAS) bridges the two corpora to generate low-dimensional, informative, and interpretable coordinates. Figure \ref{fig:workflow} summarizes the architecture.

RAS uses two corpora with different jobs. Let $\mathcal{A}$ be the \emph{anchor corpus}: it provides organized material from which RAS estimates semantic anchors. Let $\mathcal{R}$ be the \emph{response corpus}: it provides the naturally occurring text to be scored. RAS learns the anchors from $\mathcal{A}$ and maps each text in $\mathcal{R}$ onto them, producing topicality coordinates that play the role of comparable survey-like measures.

\subsection{Problem Setting and Assumptions}

For the formal development, let $\mathcal{R}$ denote the response corpus and $\mathcal{A}$ denote the anchor corpus. In this paper, $\mathcal{R}$ is EJMR and $\mathcal{A}$ consists of elite-journal publications. Text objects are embedded by
\begin{equation}
\phi:T_{type}\rightarrow\mathbb{R}^{d},\qquad type \in \{r,a\}
\end{equation}
where $r$ means this text $T_{r,i}$ comes from response corpus $\mathcal{R}$, and $a$ for anchor corpus $\mathcal{A}$, $\phi$ is a pre-trained embedding model that converts inputs into high-dimensional vector. So response-corpus text $T_{r,i}\in\mathcal{R}$ has embedding $r_i=\phi(T_{r,i})$ and anchor-corpus text $T_{a,j}\in\mathcal{A}$ has embedding $a_j=\phi(T_{a,j})$ with dimension $d$.

The researcher's domain of interest, called the anchor-corpus space below, is represented by a $k$-dimensional anchor-corpus topic basis
\begin{equation}
V_a=[v_{a,1},\dots,v_{a,k}]\in\mathbb{R}^{d\times k},
\qquad k\ll d.
\end{equation}
In this paper, these directions are the research topics used to organize professional discussion. Each vector $v_{a,i}$ can stand for one topic the researchers are interested in, such as wages and unemployment. The anchor corpus is the sample used to estimate this target basis. We represent each response-corpus embedding as
\begin{equation}
r_i=V_a b_i+V_\perp c_i+u_i,
\qquad
V_\perp^\top V_a=0,
\end{equation}
where $V_\perp\in\mathbb{R}^{d\times(d-k)}$ contains basis vectors orthogonal to $V_a$, and $u_i$ is an embedding or sampling perturbation with $\mathbb{E}[u_i\mid b_i,c_i]=0$ and finite second moments. 

This setting is useful because in reality there often appears both relevant topics and irrelavant topics in the same text. Making the basis $V_a$ and $V_\perp$ orthogonal not only keeps as much as possible the information of $V_a$ by splitting the orthogonal or irrelavant information $V_\perp$, but also is consistent with common orthogonal assumptions in linear algebra. For example, if we are studying macro economics, then topics like sports are just in $V_\perp$.

Let $\theta$ be the parameter we are concerned about. The theoretical target can be written as $\theta_0=f(V_a)$, a functional of the anchor-corpus topic basis.
\begin{equation}
f_0(\theta)=\mathbb{E}[\mathcal{L}(\theta;b_i,y_i)],
\qquad
\theta_0\in\arg\min_{\theta\in\Theta} f_0(\theta),
\end{equation}
where $\mathcal{L}$ is a generic loss or estimating criterion evaluated on the reference-anchored coordinates of response-corpus observations. Also, we propose the following assumption:

\paragraph{Assumption (anchor-corpus estimability).} The anchor corpus contains enough information to estimate the population topic basis $V_a$. Anchor-corpus texts are generated by
\begin{equation}
a_j=V_a d_j+\epsilon_j.
\end{equation}
For each topic direction $\ell$, there is an observed or constructible anchor-corpus subset $\mathcal{J}_\ell$ whose average coordinate loads primarily on that basis vector:
\begin{equation}
\mathbb{E}[d_j\mid j\in\mathcal{J}_\ell]=\lambda_\ell e_\ell,
\qquad
\mathbb{E}[\epsilon_j\mid j\in\mathcal{J}_\ell]=0,
\qquad
\mathbb{E}[\|\epsilon_j\|_2^2\mid j\in\mathcal{J}_\ell]\leq \sigma_{r\ell}^2<\infty,
\end{equation}
where $e_\ell$ is the $\ell$th coordinate vector and $\lambda_\ell>0$. Thus the population topic center $\mu_\ell=\mathbb{E}[q_j\mid j\in\mathcal{J}_\ell]$ is proportional to $v_{a,\ell}$, and the normalized center recovers the population direction. 

\subsection{Reference-Anchored Semantic Reparameterization}

RAS changes the representation used inside the criterion. The ideal criterion $f_0(\theta)$ is written in terms of the $k$-dimensional reference-anchored content of response-corpus observations, but the corresponding topic directions are not observed. So for each topic $\ell$, we select a topic-relevant subset $\mathcal{A}_\ell\subseteq\mathcal{A}$. This subset could come from the clustering process of the embedding vectors. Let $|\mathcal{A}_\ell|$ denote the number of elements in set $\mathcal{A}_\ell$. The estimated topic direction is the average\footnote{If necessary, we can also use weighted average with positive weights $w_{i,j}$ that sum to one for each topic.} of the anchor-corpus embeddings in that subset:
\begin{equation}
\widehat{v}_\ell
=\frac{\sum_{a_j\in\mathcal{A}_\ell}a_j}
{|\mathcal{A}_\ell|}
\end{equation}
The subsets and weights can be obtained from clustering, retrieval, or another topic-construction procedure. Under Assumption, this weighted average estimates the corresponding column of the anchor-corpus topic basis. Collect the estimated directions as $\widehat{V}=[\widehat{v}_1,\dots,\widehat{v}_k]$. For each response-corpus observation $i\in\mathcal{R}$, RAS constructs the interpretable coordinates
\begin{equation}
s_i=\widehat{V}^{\top}r_i\in\mathbb{R}^{k}.
\end{equation}
The component $s_{i\ell}=\widehat{v}_\ell^\top r_i$ measures the alignment of response-corpus observation $i$ with anchor-corpus topic $\ell$. Thus RAS turns the original high-dimensional embedding into the $k$ topic coordinates used to estimate $\theta$. Under the target-basis representation,
\begin{equation}
r_i=V_a b_i+V_\perp c_i+u_i,
\qquad
s_i=\widehat{V}^{\top}r_i
=A b_i+\widehat{V}^{\top}V_\perp c_i+\widehat{V}^{\top}u_i,
\qquad
A=\widehat{V}^{\top}V_a.
\end{equation}
Assumption implies $A\to V_a^\top V_a$, while $\widehat{V}^{\top}V_\perp\to V_a^\top V_\perp=0$, $\widehat{V}^{\top}u_i \to 0$. Hence $A$ is invertible and
\begin{equation}
b_i=A^{-1}s_i-A^{-1}\widehat{V}^{\top}V_\perp c_i-A^{-1}\widehat{V}^{\top}u_i \to A^{-1}s_i.
\end{equation}

RAS preserves the information carried by $V_a$ up to the projected embedding perturbation. The resulting $s_i$ provide interpretable coordinates for the target topic basis.
The main advantage of RAS is interpretability. It replaces the high-dimensional embedding $x_i\in\mathbb{R}^d$ with
\begin{equation}
s_i=(s_{i1},\dots,s_{ik})^\top,
\qquad
s_{i\ell}=\widehat{v}_\ell^\top x_i,
\end{equation}
so that each coordinate is anchored to an anchor-corpus topic. The resulting parameterization of $\theta$ is low-dimensional and directly interpretable, whereas a parameter attached to the raw coordinates of $r_i$ has no comparable topic meaning. 

The representation is also invariant to a common orthogonal rotation $H$, which means if we use anther embedding model $\phi_{new}$, as long as the relative similarity is constant\footnote{This is often satisfied by modern embedding models, since their training target is the semantic similarity of real world}, the RAS result is constant:
\begin{equation}
(H\widehat{V})^\top(Hr_i)=\widehat{V}^\top r_i=s_i.
\end{equation}

\subsection{TaDaS Output: Survey-Like Measurement Dataset}

Reference-Anchored Semantic Reparameterization (RAS) estimates the $k$ anchor-corpus topic directions from $\mathcal{A}$ and re-expresses response-corpus text in those coordinates.

\begin{equation}
s_i=\widehat{V}^{\top}r_i.
\end{equation}
The resulting survey-like estimating dataset structure is $\{s_i,Z_i:i\in\mathcal{R}\}$, with optional outcomes $y_i$ appended only when the application requires them. In this structure, $s_i$ is the parameter we construct from high-dimensional embedding vector using the method RAS, $Z_i$ is the covariates we can obtain. Then the final survey-like structure $\mathcal{D}_{TaDaS}$ could be seen as:
\begin{equation}
\mathcal{D}_{TaDaS}
=
\{y_i,s_i,Z_i:i\in\mathcal{R}\}.
\end{equation}

This construction separates measurement from outcome scoring. TaDaS constructs the topic-anchored variables used to estimate and interpret $\theta$. Attitude scoring, human coding, metadata, or observed behavior supply $y_i$ only in applications that need an outcome.

In the present application, the anchor corpus consists of elite-journal publications, whose topic structure is summarized by the basis $\widehat{V}$. The EJMR replies form the response corpus. RAS maps those replies onto the semantic anchors to generate topicality measures, including $\mathrm{Topicality}_{AI}$ and the other topic measures. A topicality measure is an observation's semantic proximity to an anchor-corpus topic center; it is not a measure of individual treatment, technology adoption, or causal exposure. The downstream attitude layer then adds reply-level outcomes such as openness, negativity, poisonousness, and curiosity. Figure \ref{fig:workflow} presents the full framework.

To make the progress more clear, consider an EJMR reply that asks whether ChatGPT can help economists write code for empirical research. First, the anchor corpus supplies a set of AI-related publications---for example, studies of AI's effects on labor markets, firms' production processes, and research practice. Although these studies examine distinct substantive settings, they are organized around AI; their shared semantic content allows their embeddings to be summarized as an estimate of the AI topic embedding. RAS then embeds the reply and projects it onto the full set of estimated topic anchors. A relatively large $s_{i,AI}$ indicates that this reply is semantically close to the AI research topic; its other coordinates record any simultaneous proximity to other anchor-corpus topics. In this sense, RAS turns the reply into a comparable response to the implicit question, ``How much does this text concern AI-related research?'' If the application also scores the reply's openness or negativity, that score is appended as an outcome only after topicality has been measured. Thus, an AI-topical reply can express either openness or resistance: the topicality coordinate identifies what the reply concerns, while the outcome records how it is expressed.

\subsection{Potential Applications}

TaDaS is portable when a \textbf{anchor corpus} supplies an interpretable semantic basis and a \textbf{response corpus} contains the naturally occurring text to be measured. The anchor corpus need not be a survey: publications, classifications, standards, financial records, and course structures can all provide stable semantic coordinates. The response corpus can consist of posts, reviews, news, transcripts, or other unstructured observations. In each application, the design should make explicit the two corpora and the decision-relevant outcome that the resulting coordinates will inform. For application $a$, write
\begin{equation}
\widehat{V}^{(a)}=\Psi_A(\mathcal{A}^{(a)}),
\qquad
s_i^{(a)}=(\widehat{V}^{(a)})^\top x_i.
\end{equation}
Here, $\mathcal{A}^{(a)}$ is the anchor corpus and $\mathcal{R}^{(a)}$ is the response corpus. RAS represents each response observation in interpretable, reference-anchored coordinates.

\paragraph{1. Stock and investment.} The anchor corpus could comprise financial statements, firm classifications, analyst taxonomies, product categories, and historical asset characteristics. The response corpus could comprise financial news, earnings-call transcripts, investor discussions, and analyst commentary. RAS could score that language along dimensions such as profitability, innovation, risk, and competition to inform decisions about portfolio monitoring, research prioritization, or the interpretation of market sentiment; the resulting measures remain descriptive inputs rather than mechanically identified trading signals.

\paragraph{2. Data science, e-commerce, marketing, decision science, and politics.} The anchor corpus could provide product taxonomies, customer attributes, decision categories, policy classifications, or survey instruments. The response corpus could include customer reviews, product descriptions, public comments, social-media discussion, and employee or user feedback. The resulting coordinates could inform product positioning, service redesign, demand analysis, or the interpretation of policy attitudes by showing which structured dimensions organize the noisy text.

\paragraph{3. Education economics.} The anchor corpus could contain course catalogs, syllabi, learning objectives, program classifications, instructor attributes, and curriculum taxonomies. The response corpus could contain online course reviews, MBA-course discussions, student feedback, alumni comments, and education-policy debate. TaDaS could construct coordinates for perceived teaching quality, workload, skill relevance, career value, and engagement to inform curriculum design, program evaluation, or student-support decisions while retaining the original text for validation and qualitative interpretation.

Across these applications, the common contribution is to make unstructured text measurable without imposing a researcher-written questionnaire. TaDaS uses an anchor corpus to define interpretable semantic anchors, uses RAS to map naturally occurring text into topicality coordinates, and attaches outcomes only when an application requires them. The resulting variables support cautious descriptive analysis and, where validated, downstream social-science estimation.
\section{Related Literature}
\label{sec:literature}

\subsection{Tech Background}

Transformer models such as \citet{vaswani2017}, \citet{brown2020}, and \citet{ouyang2022} provide the broader language-modeling foundation for modern representation learning and large language models. The Transformer architecture makes it possible to represent contextual relations across a sequence, while LLMs extend this foundation to broad language understanding and generation.

Embedding methods shift the measurement problem from word co-occurrence to semantic proximity. \citet{reimers2019} show that sentence-transformer representations can make semantic similarity practical at scale.

The measurement strategy also builds on a topic-modeling tradition that treats large text collections as structured empirical objects. \citet{blei2003} introduced latent Dirichlet allocation as a probabilistic way to recover interpretable topics from high-dimensional word counts, and \citet{roberts2014} extended topic modeling toward open-ended survey responses and document-level covariates. This line of work matters for the present paper because it shows how text can be converted from raw language into a lower-dimensional empirical representation rather than read only as individual documents. The paper's topic construction further draws on dimensionality reduction and density-based clustering tools, especially UMAP \citep{mcinnes2018umap}, HDBSCAN \citep{campello2013hdbscan,mcinnes2017hdbscan}, and BERTopic-style c-TF-IDF topic summaries \citep{grootendorst2022bertopic}.

\subsection{NLP for Social Sciences}

Natural-language processing and text-as-data methods have become central tools in economics and related social sciences. Methodological work emphasizes the promise of scalable measurement while stressing validation and explicit design choices, and recent computational-linguistics research extends these methods to corporate-generated text \citep{gentzkow2019,grimmer2013,hassan2025}. A broad set of applications uses text to measure policy uncertainty, financial language, product-market structure, and business-cycle conditions \citep{baker2016,loughran2011,hoberg2016,xiu2024}.

Recent work combines richer embeddings, topic models, sentiment measures, and LLMs to study financial information and organizational text \citep{fan2024,siano2025,li2024,lin2024}. The broader AI literature further examines exposure, productivity, work organization, and scientific production as AI capabilities develop \citep{autor2015,felten2021,acemoglu2022,eloundou2023,noy2023,brynjolfsson2025,kanazawa2025,hao2025,pramanick2026}. Together, these studies motivate using NLP to measure how economic and professional environments respond to technological change.

The cross-corpus use of semantic structure also connects to studies that use topic movement to measure institutional or intellectual change. \citet{locatelli2023} use topic shifts to study politicization in public communication, \citet{goldsmithpinkham2024} traces the credibility revolution across fields, and \citet{bello2025} uses research similarity to study gendered academic patterns. \citet{pramanick2026} similarly studies how large language models reshape scholarly landscapes across fields. This paper follows these studies in treating semantic movement as substantively meaningful, but it uses semantic anchors estimated from an anchor corpus to organize a response corpus.
\subsection{NLP for Anonymous Professional Discourse}

The EJMR--AI illustration relates to research on anonymous professional discourse. \citet{wu2018} shows that gendered language on EJMR reflects broader asymmetries in the economics profession, and \citet{ederer2025} study attention and abuse in anonymous economics discourse. These studies establish that anonymous professional text can document hierarchy, exclusion, and conflict while also demanding careful interpretation.

Work outside economics clarifies why this type of response corpus differs from formal survey responses or public professional statements. \citet{suler2004} describes the online disinhibition effect, and \citet{cheng2017} show that antisocial behavior in online discussions depends partly on context and prior interaction. \citet{alberti2024} documents toxic communication around the accounting academic job market, while \citet{flores2023} distinguishes passive and proactive online behavior modes. Together, these findings make anonymous discourse a stringent test of a measurement framework: its language is naturally occurring and revealing, but also noisy and context dependent.

For TaDaS, EJMR is therefore a response corpus rather than a stand-in for the profession. Its anonymity and selection require restraint about whose beliefs the archive reflects. The illustration tests whether transparent reference anchoring can organize heterogeneous professional text while preserving those limitations.

This interpretation follows the discourse literature: anonymous professional archives can reveal status asymmetries and abuse that formal communication may mute, while the same setting can amplify disinhibition and context-dependent antisocial behavior \citep{wu2018,ederer2025,suler2004,cheng2017}. TaDaS addresses the measurement problem by making the semantic reference frame explicit; it does not remove the need to delimit the archive or interpret its descriptive patterns cautiously.

\subsection{Contributions}

Relative to these literatures, the paper contributes a method for disciplined measurement from archival text; EJMR--AI is the illustrative application.

First, TaDaS provides a measurement architecture. It separates the anchor corpus that defines semantic content from the response corpus that supplies naturally occurring language, then turns their relationship into reusable topicality coordinates. This extends text-as-data work beyond document classification toward a general design for recovering comparable measures when direct surveys are costly, infrequent, or vulnerable to self-presentation.

Second, RAS makes semantic anchoring transparent. Anchor construction, corpus roles, projection, and downstream outcomes are explicit rather than implicit in a black-box score. Researchers can therefore inspect how an anchor corpus supplies the dimensions used to measure response text, validate those dimensions against domain knowledge, and distinguish measurement from any later substantive analysis.

Third, the EJMR--AI illustration demonstrates the framework in noisy professional discourse. It combines publication-derived semantic anchors, response-corpus topicality, and downstream attitude scoring to characterize patterns within the archive. The illustration shows how TaDaS can organize disagreement, status concerns, adaptation, and exclusion in naturally occurring text without treating the archive as a population sample or assigning causal meaning to its descriptive associations.

\section{Illustrative Application: Economists' Responses to AI}
\label{sec:data}

\subsection{Anchor-Corpus and Topic Anchors Construction}

This demonstration asks how responses to artificial intelligence (AI) are organized in an archive of professional discourse. It uses elite economics and finance journal publications as the anchor corpus and EJMR replies as the response corpus. The response corpus records repeated, naturally occurring professional discourse, but it is platform-specific and nonrepresentative; accordingly, the demonstration describes patterns within the archive rather than among economists as a population. The anchor corpus contains 53,585 papers drawn from elite economics and finance journals closely aligned with the ABS 4* tier. These publications provide the semantic material used to estimate the topic structure. This use of publications as reference material is consistent with classic accounts of scientific recognition and the diffusion of ideas through scholarly communities \citep{merton1968,bourdieu1975,crane1972}.

\begin{figure}[H]
    \centering
    \begin{minipage}[t]{0.48\textwidth}
        \centering
        \includegraphics[width=\linewidth]{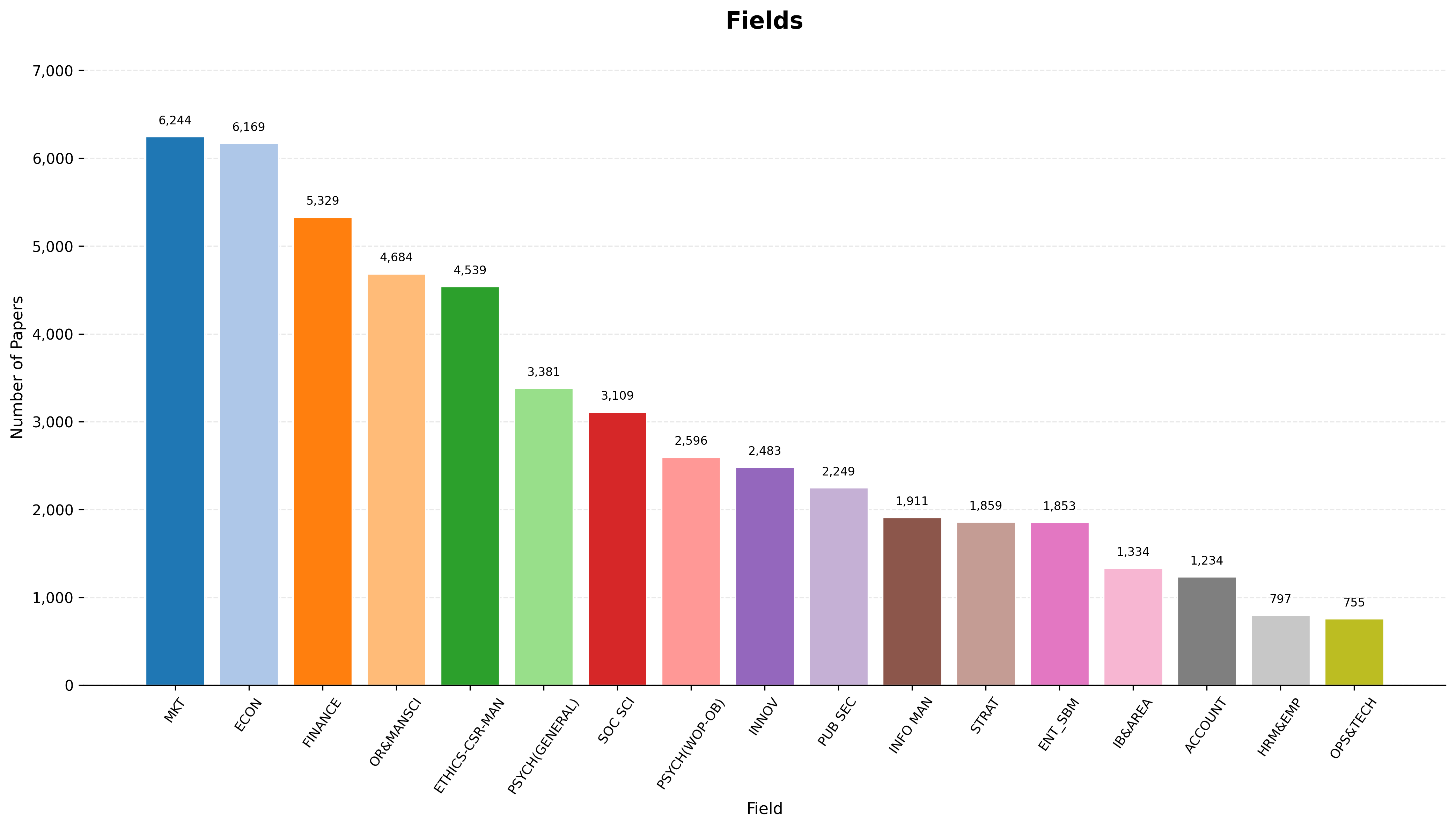}
        \captionof{figure}{Anchor-Corpus Field Distribution}
        \label{fig:field_distribution}
    \end{minipage}\hfill
    \begin{minipage}[t]{0.48\textwidth}
        \centering
        \includegraphics[width=\linewidth]{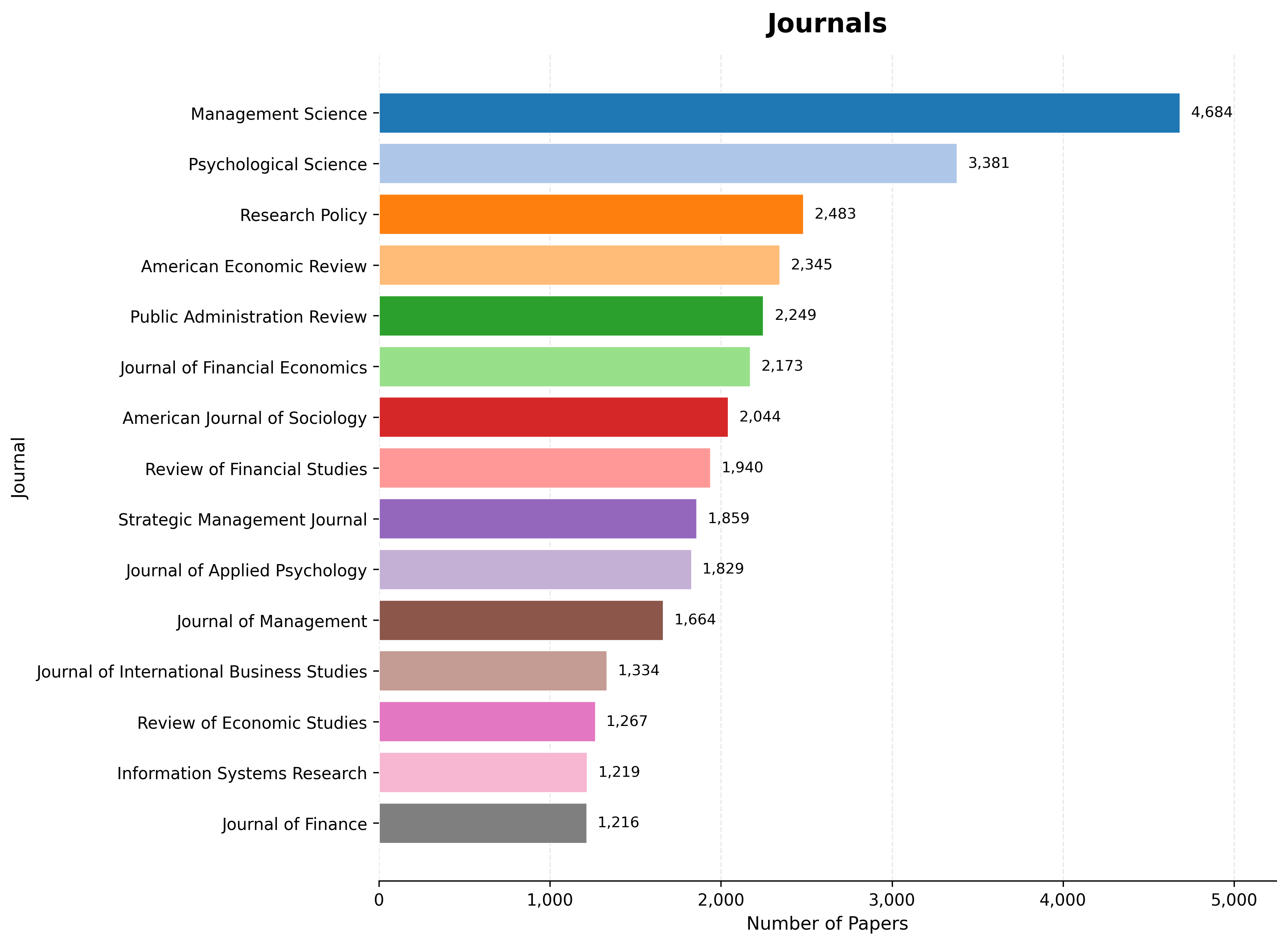}
        \captionof{figure}{Anchor-Corpus Journal Distribution}
        \label{fig:journal_distribution}
    \end{minipage}
\end{figure}

\begin{figure}[H]
    \centering
    \includegraphics[width=\textwidth]{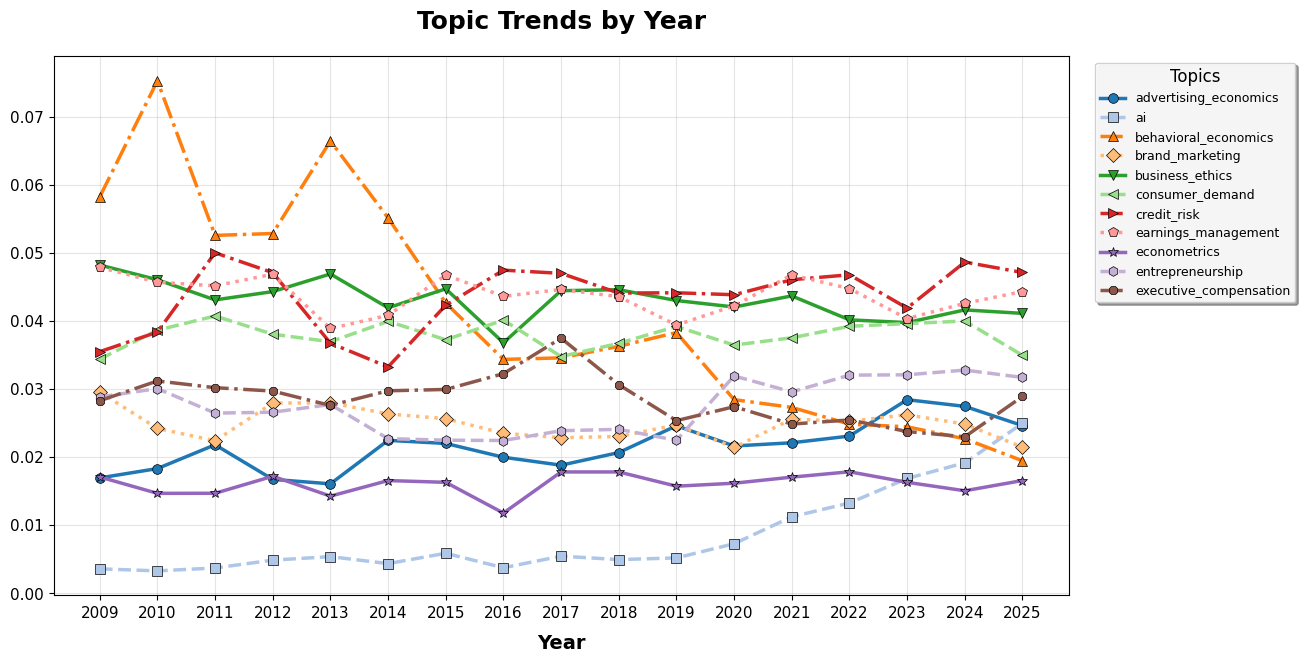}
    \caption{Annual Anchor-Corpus Research Trends}
    \label{fig:proxy_trend}
\end{figure}

Each anchor-corpus publication record contributes title and abstract text. These texts are embedded with All-RoBERTa Large v1 and grouped with UMAP and HDBSCAN; cluster-level keywords are then recovered with class-based TF-IDF. The resulting inventory covers AI, econometrics, health economics, family business, labor and wages, gender economics, personnel economics, and so on. In this demonstration, TaDaS uses these topic centers to map EJMR replies into interpretable topicality measures such as $\mathrm{Topicality}\_{AI}$. Aggregating the intensity of each anchor-corpus topic by publication year also produces annual anchor-corpus research trends. Figure \ref{fig:proxy_trend} reports selected series and shows how these trends develop over time.

To make the anchor-corpus construction transparent, Table \ref{tab:paper_topic_keywords} reports the full topic inventory together with representative high-weight keywords, while Appendix Table \ref{tab:paper_topic_cases} reports illustrative paper titles associated with each topic. Appendix Section \ref{app:topic_clusters} provides further details on the embedding, clustering, and topic-center construction.

\begingroup
\scriptsize
\setlength{\LTcapwidth}{\textwidth}
\begin{longtable}{p{0.22\textwidth}p{0.06\textwidth}p{0.62\textwidth}}
\caption{Anchor-Corpus Topics and Representative Keywords}\label{tab:paper_topic_keywords}\\
\toprule
Topic & Papers & Representative Keywords \\
\midrule
\endfirsthead
\toprule
Topic & Papers & Representative Keywords \\
\midrule
\endhead
Artificial Intelligence (AI) & 495 & artificial intelligence; intelligence ai; ai; human machine; machine learning \\
Econometrics & 618 & estimators; estimator; estimating; estimated; inference \\
Health Economics & 543 & patients; hospitals; hospital; health insurance; healthcare \\
Family Business & 458 & family firms; family firm; families; family; owned firms \\
Labor and Wages & 1,549 & wages; wage; minimum wage; unemployment; wage inequality \\
Gender Economics & 1,400 & gender diversity; gender bias; gender equality; gender gap; gender differences \\
Personnel Economics & 904 & human resource; human resources; hr; organizational performance\\
Public Governance & 2,160 & administration research; bureaucracy; governance; bureaucracies\\
Firm Strategy & 3,153 & organization studies; management research; strategy research \\
Behavioral Economics & 2,183 & working memory; perceptual; visual; perception; memory \\
Venture Capital & 1,278 & venture capitalists; venture capital; investors; capital firms; startups \\
Entrepreneurship & 2,105 & entrepreneurship literature; entrepreneurship; entrepreneurial activity \\
Supply Chain & 741 & supply chains; value chain; supply chain; suppliers; supplier \\
Game Theory & 1,141 & nash equilibrium; payoffs; equilibrium outcome; cooperation; equilibria \\
Mergers Acquisitions & 746 & mergers acquisitions; acquisitions; merger acquisition; acquisition; mergers \\
Executive Compensation & 1,826 & corporate governance; ceo compensation; ceo turnover; executive compensation \\
Online Reviews & 758 & product reviews; online reviews; reviews; sentiment; rating systems \\
Platform Economics & 768 & platform owners; digital platforms; digital platform; platforms; revenue sharing \\
Innovation and Patents & 2,661 & patents; patenting; innovation performance; patent; open innovation \\
International Business & 2,157 & multinational enterprises; international business; emerging markets \\
Credit Risk & 3,005 & borrowers; credit risk; lending; default risk; credit default \\
Business Ethics & 3,336 & ethical leadership; organizational justice; unethical behavior; moral \\
Team Productivity & 1,287 & leadership; organizational citizenship; team performance; teamwork \\
Advertising Economics & 630 & advertisements; advertisement; ad; ads; consumer search \\
Retail Pricing & 1,165 & retailers; price competition; wholesale; dynamic pricing; inventory \\
Racial Inequality & 1,186 & racial disparities; police; blacks; officers; policing \\
Intergenerational Mobility & 1,896 & intergenerational mobility; income inequality; social mobility \\
Risk Preferences & 3,926 & risk attitudes; ambiguity aversion; risk preferences; prospect theory \\
Consumer Demand & 1,802 & consumer psychology; purchase decisions; price discounts; discounts; discount \\
Brand Marketing & 3,661 & brands; consumers perceive; consumer psychology; purchase intentions; brand \\
Hedge Funds & 1,666 & hedge fund; hedge funds; fund managers; fund performance; mutual funds \\
Earnings Management & 2,381 & earnings forecasts; earnings management; earnings; financial reporting \\
\bottomrule
\end{longtable}
\endgroup

\subsection{Mapping Response-Corpus to Topic Anchors}

With the anchor-corpus topic system defined, this demonstration maps the EJMR response corpus into the same semantic space. EJMR is an anonymous professional discussion environment that combines serious research discussion, labor-market anxiety, status signaling, methodological debate, and conflict. This mixture makes EJMR noisy, but it also makes it useful for this demonstration: the platform preserves naturally occurring professional reactions that would often be softened in a direct survey \citep{bertrand2001,tourangeau2007}. Online Appendix Figure \ref{fig:ejmr_frontpage} shows the discussion environment from which the response corpus is drawn.

Table \ref{tab:motivating_quotes} illustrates why this response corpus is useful for measuring professional reactions to AI. The same thread contains both gatekeeping responses that defend existing training standards and open responses that treat generative AI, reinforcement learning, and NLP as part of the future skill set for economists.

\begin{table}[H]
	\centering
\caption{Illustrative EJMR Responses to AI}
	\label{tab:motivating_quotes}
	\small
	\setlength{\tabcolsep}{6pt}
	\renewcommand{\arraystretch}{1.5}
	\begin{tabularx}{\textwidth}{@{}>{\raggedright\arraybackslash}X | >{\raggedright\arraybackslash}X@{}}
		\toprule
		\multicolumn{2}{@{}l}{\href{https://www.econjobrumors.com/topic/should-the-econ-grad-core-include-generative-ai-reinforcement-learning-nlp}{\textbf{Thread title}: Should the econ grad core include Generative AI, Reinforcement Learning, NLP? (2023)}} \\
		\midrule
		\textbf{Gatekeeping views} & \textbf{Open views} \\
		\midrule
		\textit{``Do you even know what reinforcement learning is?''} &
		\textit{``Seems like these are the future and econ hasn't changed since 1950.''} \\
		\midrule
		``RL is basically approximate dynamic programming---a staple of the curriculum since the 80s.'' &
		``To get hired in 2023 without an academic job, we're stuck taking Coursera and Udemy classes...'' \\
		\midrule
		``Making it `core' means qualifying exams on ML, which is nonsense.'' &
		``...this is after 5 years post-undergrad in Econ. It's time to adapt.'' \\
		\bottomrule
	\end{tabularx}
	\caption*{\footnotesize Notes: Selected posts from the same EJMR thread.}
\end{table}

The broader raw EJMR response archive contains 7,989,637 posts. This corpus contains topic areas far from research or professional academic activity. Alongside threads about methods, journals, research trends, and academic work, the platform includes everyday conversation about entertainment, food, exercise, cars, and other non-research subjects. representative examples appear in Figure \ref{fig:ejmr_nonresearch_examples}.

\begin{figure}[H]
	\centering
	\includegraphics[width=0.88\textwidth]{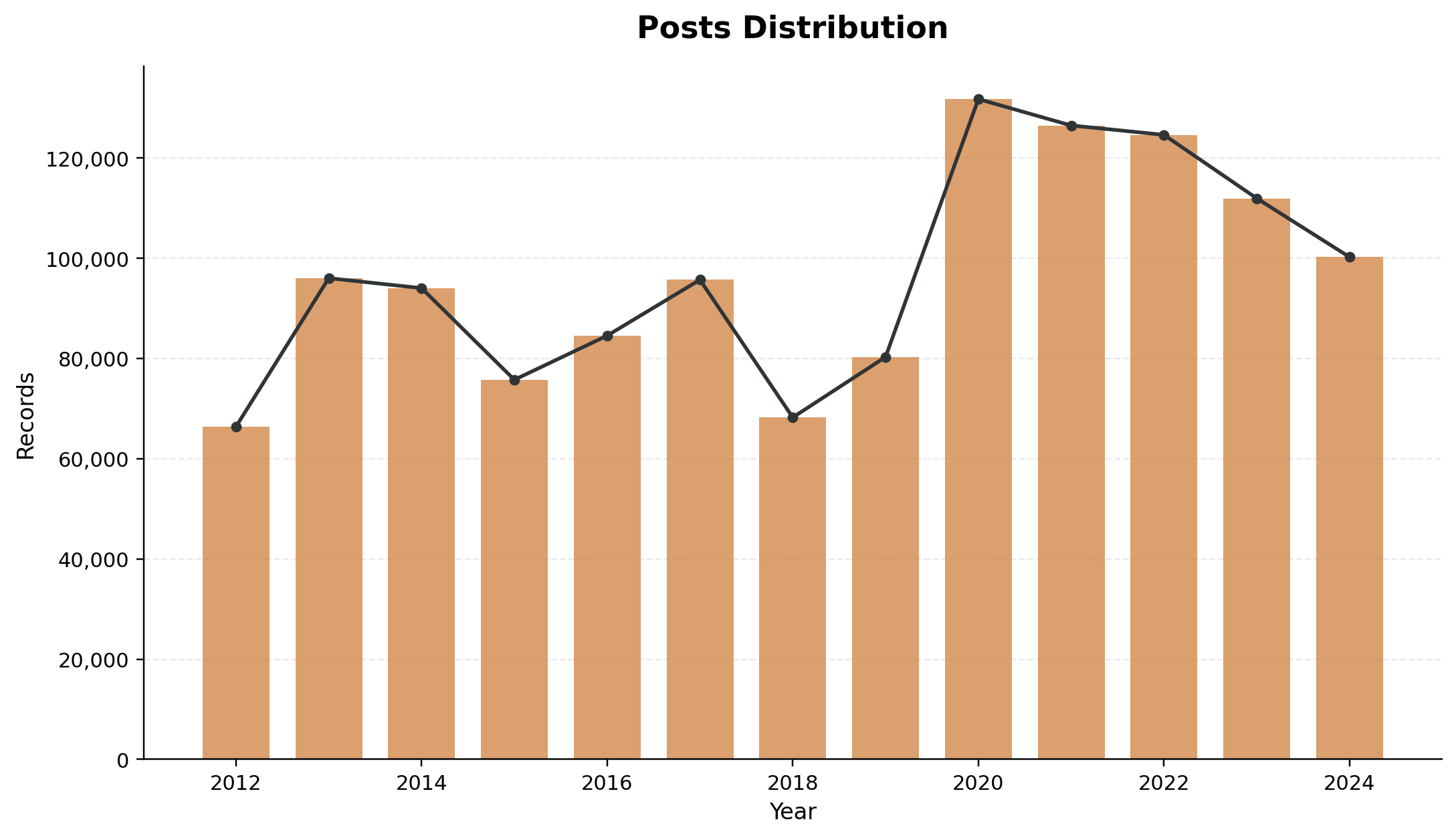}
	\caption{EJMR Response-Corpus Posts Over Time}
	\label{fig:ejmr_topic_distribution}
\end{figure}

To focus on the topics we're interested in, we use self-clustering to identify academic related topics and their corresponding posts as our data source. The EJMR analysis data contain a regression estimation sample of 1,255,942 observations, ranging from year 2012 to 2024. Figure \ref{fig:ejmr_topic_distribution} reports the annual distribution used in this demonstration.

In this demonstration, TaDaS measures EJMR response-corpus topicality by passing reply embeddings through the same encoder used on the anchor corpus and then mapping each response observation to the anchor-corpus topic centers.

The paper therefore identifies topicality semantically rather than through a narrow keyword list. A post has high AI topicality when its embedding is close to the AI topic center, even when the text uses related language such as machine learning, coding, automation, software tools, or changes in research skills.

The following table reports our main RAS results: the reference-anchored semantic coordinates constructed from the anchor corpus and used to measure response-corpus topicality.

\begin{center}
\scriptsize
\setlength{\tabcolsep}{4pt}
\begin{longtable}{lrrrrr}
\caption{Descriptive Statistics for Response-Corpus Topicality Measures}\label{tab:descriptive_topics_appendix}\\
\toprule
Variable & N & Mean & Std. Dev. & Min & Max \\
\midrule
\endfirsthead
\multicolumn{6}{l}{\tablename\ \thetable{} -- continued from previous page}\\
\toprule
Variable & N & Mean & Std. Dev. & Min & Max \\
\midrule
\endhead
\midrule
\multicolumn{6}{r}{Continued on next page}\\
\endfoot
\bottomrule
\endlastfoot
\addlinespace
\multicolumn{6}{l}{\textit{Response-Corpus Topicality Measures ($\mathrm{Topicality}_*$)}} \\
Artificial Intelligence (AI) & 1,255,942 & 0.1069 & 0.1010 & -0.1747 & 0.6756 \\
Econometrics & 1,255,942 & 0.1109 & 0.1052 & -0.1575 & 0.7293 \\
Health Economics & 1,255,942 & 0.0631 & 0.0772 & -0.1713 & 0.6961 \\
Family Business & 1,255,942 & 0.0625 & 0.0760 & -0.1623 & 0.6697 \\
Labor Wages & 1,255,942 & 0.0870 & 0.0960 & -0.1567 & 0.6965 \\
Gender Economics & 1,255,942 & 0.0885 & 0.0866 & -0.1831 & 0.6645 \\
Personnel Economics & 1,255,942 & 0.0573 & 0.0811 & -0.1880 & 0.6176 \\
Public Governance & 1,255,942 & 0.0753 & 0.0808 & -0.1664 & 0.5821 \\
Firm Strategy & 1,255,942 & 0.1127 & 0.1009 & -0.1644 & 0.6777 \\
Behavioral Economics & 1,255,942 & 0.1300 & 0.0801 & -0.1550 & 0.4953 \\
Venture Capital & 1,255,942 & 0.0875 & 0.0984 & -0.1653 & 0.7507 \\
Entrepreneurship & 1,255,942 & 0.0723 & 0.0802 & -0.1615 & 0.7306 \\
Supply Chain Economics & 1,255,942 & 0.0454 & 0.0815 & -0.1848 & 0.5419 \\
Game Theory & 1,255,942 & 0.0739 & 0.0908 & -0.1804 & 0.7145 \\
Mergers Acquisitions & 1,255,942 & 0.0723 & 0.0933 & -0.1765 & 0.7974 \\
Executive Compensation & 1,255,942 & 0.0780 & 0.0924 & -0.1916 & 0.7405 \\
Online Reviews & 1,255,942 & 0.0896 & 0.0848 & -0.1801 & 0.6071 \\
Platform Economics & 1,255,942 & 0.0806 & 0.0909 & -0.1628 & 0.7064 \\
Innovation Patents & 1,255,942 & 0.0830 & 0.0901 & -0.1900 & 0.7821 \\
International Business & 1,255,942 & 0.0730 & 0.0887 & -0.1655 & 0.5930 \\
Credit Risk & 1,255,942 & 0.0755 & 0.0983 & -0.1772 & 0.6832 \\
Business Ethics & 1,255,942 & 0.0739 & 0.0853 & -0.1698 & 0.5989 \\
Team Productivity & 1,255,942 & 0.0528 & 0.0727 & -0.1847 & 0.6338 \\
Advertising Economics & 1,255,942 & 0.0818 & 0.0831 & -0.1590 & 0.6087 \\
Retail Pricing & 1,255,942 & 0.0539 & 0.0913 & -0.2004 & 0.6381 \\
Racial Inequality & 1,255,942 & 0.0890 & 0.0856 & -0.1742 & 0.6715 \\
Intergenerational Mobility & 1,255,942 & 0.1177 & 0.1017 & -0.1690 & 0.7300 \\
Risk Preferences & 1,255,942 & 0.1208 & 0.1035 & -0.1716 & 0.7197 \\
Consumer Demand & 1,255,942 & 0.0864 & 0.0930 & -0.1549 & 0.7267 \\
Brand Marketing & 1,255,942 & 0.0871 & 0.0825 & -0.1614 & 0.6964 \\
Hedge Funds & 1,255,942 & 0.0951 & 0.1105 & -0.1509 & 0.7277 \\
Earnings Management & 1,255,942 & 0.1098 & 0.1178 & -0.1889 & 0.7227 \\
\end{longtable}
\end{center}

\begin{table}[H]
    \centering
    \caption{Sample Summary}
    \label{tab:sample_summary}
    \begin{tabular}{p{0.57\textwidth}r}
        \toprule
        Item & Count \\
        \midrule
        EJMR Response-Corpus Analysis Sample & 1,255,942 \\
        Raw Anchor Corpus Sample& 53,585 \\
        Anchor-Corpus Topicality Centroids & $32$ topics \\
        EJMR Response-Corpus Sample Period & $2012$--$2024$ \\
        \bottomrule
    \end{tabular}
\end{table}

The sample summary shows that this demonstration is large along both margins: the EJMR response corpus is large enough to support high-dimensional text measurement, and the anchor corpus is broad enough to construct a stable research-topic basis.
In this way, each EJMR reply is assigned two distinct sets of measures.  Reference-Anchored Semantic
Reparameterization (RAS) maps the reply into the anchor-corpus topic system, yielding a topicality
score for its semantic proximity to each disscusion from the EJMR archive. 

\subsection{Sentiment and Attitude Scoring}

 The response-scoring stage measures the attitude expressed in the replies for different research topics. At this stage, we are interested in understanding the emotional tone and sentiment of each reply. The four outcomes are openness, curiosity, negative tone, and poisonousness.

We implement the response-scoring stage with the instruction-tuned decoder model Gemma 3 12B IT (\texttt{gemma-3-12b-it}) \citep{gemmateam2025}. Using an instruction-following model as a structured text annotator is consistent with recent evidence that large language models can support social-science annotation at scale \citep{gilardi2023,ziems2024}. For each prompt post $Q_j$ and reply $A_j$, the model returns a vector of four attitude scores, each bounded between $0$ and $1$. The scoring rubric defines $0.5$ as no clear signal; values above or below that midpoint indicate stronger evidence in the corresponding direction. The complete scoring prompt appears in Appendix \ref{app:sentiment_prompt}, and Appendix Table \ref{tab:scoring_audit_cases} presents illustrative reply-level checks.

\begin{table}[H]
    \centering
    \caption{Descriptive Statistics for Core Variables}
    \label{tab:descriptive_core}
    \begin{tabular}{lrrrrr}
        \toprule
        Variable & N & Mean & Std. Dev. & Min & Max \\
        \midrule
        Openness & 1,255,942 & 0.4556 & 0.2247 & 0.0000 & 1.0000 \\
        Curiosity & 1,255,942 & 0.4081 & 0.2196 & 0.0000 & 1.0000 \\
        Negative & 1,255,942 & 0.5540 & 0.2353 & 0.0000 & 1.0000 \\
        Poisonous & 1,255,942 & 0.4878 & 0.2412 & 0.0000 & 1.0000 \\

        \bottomrule
    \end{tabular}
\end{table}

The sample means are close to the rubric's neutral midpoint of $0.5$. Within this EJMR analysis sample, replies show modestly higher negative tone (0.554) and modestly lower openness (0.456) and curiosity (0.408), while poisonousness (0.488) is nearly neutral. These are descriptive properties of the archived replies, not estimates of attitudes in the broader population of economists. The regression sample contains 1,255,942 observations.

\section{Descriptive Results: AI-Related Discourse on EJMR}
\label{sec:model}

The preceding section applied Text as Data as Survey (TaDaS) to transform the unstructured EJMR reply archive into a structured analysis dataset. Each reply is now represented by topicality measures, which locate it relative to the topic anchors learned from the publication corpus, and by four reply-level attitude scores: openness, curiosity, negative tone, and poisonousness. This section uses those structured variables in descriptive regressions to characterize the association between the topic of a reply and its expressed attitude, and how those associations vary over time. The analysis describes patterns within the EJMR archive; it does not estimate the causal effects of AI or represent the attitudes of economists as a population.

\subsection{Cross-Sectional Topic Openness Comparisons}

We first use the structured topicality and attitude variables to compare AI-related replies with replies that are more closely related to other research topics in the same EJMR analysis sample. These comparisons provide descriptive associations, not causal effects of AI development on attitudes.

\begin{equation}
Y_{i,t} = \beta_0 + \sum_k \beta_k \cdot \mathrm{Topicality}_{i,k} + \sum_{y \ne 2012} \gamma_y D_{t,y} + X_i' \delta + \epsilon_{i,t},
\label{eq:cross_section}
\end{equation}

where $Y_{i,t}$ is one of the reply-level attitude outcomes, $\mathrm{Topicality}_{i,k}$ measures semantic proximity to topic $k$, and $D_{t,y}$ is an indicator for calendar year $y$. The $\gamma_y$ coefficients absorb calendar-year fixed effects, with 2012 omitted as the reference year; $X_i$ contains forum and thread-title emotion controls.

The cross-sectional results provide a descriptive benchmark for the structured variables created by TaDaS. In the full-control specification, greater AI topicality is associated with lower openness and curiosity, higher negative tone, but lower poisonousness. Table \ref{tab:main_regression_results} reports the static associations for all anchor-corpus topics in one outcome-as-column table. The contrast across topics illustrates the value of the reference-anchored system: gender-economics topicality is associated with lower openness and higher poisonousness, whereas labor-and-wages topicality displays the opposite sign pattern on most outcomes.

\begingroup
\small
\setlength{\LTcapwidth}{\textwidth}
\setlength{\tabcolsep}{2pt}
\begin{longtable}{@{}>{\raggedright\arraybackslash}p{0.39\textwidth}*{4}{>{\centering\arraybackslash}p{0.14\textwidth}}@{}}
\caption{Main Regression Results}\label{tab:main_regression_results}\\
\toprule
 & Openness & Curiosity & Negative Tone & Poisonousness \\
\midrule
\endfirsthead
\toprule
 & Openness & Curiosity & Negative Tone & Poisonousness \\
\midrule
\endhead
\addlinespace
\multicolumn{5}{l}{\textit{All Topicality Effects}}\\
Advertising Economics & -0.0684 & 0.1322*** & -0.2099*** & 0.0581 \\
Artificial Intelligence (AI) & -0.1175** & -0.1529*** & 0.1248** & -0.1067* \\
Behavioral Economics & -0.0557** & 0.0784*** & -0.0118 & 0.1920*** \\
Brand Marketing & 0.1997 & 0.1528 & -0.4141*** & -0.3972* \\
Business Ethics & -0.6927*** & -0.5901*** & 1.2873*** & 1.0801*** \\
Consumer Demand & -0.0633 & -0.2800*** & -0.0440 & -0.0006 \\
Credit Risk & -0.0292 & 0.0624* & 0.1458*** & 0.0834 \\
Earnings Management & -0.4494*** & -0.2756*** & 0.4912*** & 0.6473*** \\
Econometrics & 0.0414 & 0.0421 & -0.1228* & -0.1642*** \\
Entrepreneurship & 0.1723** & 0.1757*** & -0.0644 & 0.1040 \\
Executive Compensation & 0.1688** & 0.1806** & -0.2104** & 0.0271 \\
Family Business & -0.1863*** & -0.1695** & 0.2709*** & 0.1460* \\
Firm Strategy & 0.1485* & -0.0263 & -0.3336*** & -0.8237*** \\
Game Theory & 0.0141 & 0.1625*** & -0.1086 & 0.1323 \\
Gender Economics & -0.7090*** & -0.7405*** & 0.7773*** & 0.9401*** \\
Health Economics & -0.3513*** & -0.2161*** & 0.5325*** & 0.4789*** \\
Hedge Funds & 0.1022 & -0.0199 & 0.0027 & 0.0641 \\
Innovation and Patents & 0.0803 & 0.2212*** & 0.1902*** & 0.3406*** \\
Intergenerational Mobility & -0.0727 & -0.2159*** & -0.1383** & -0.1149 \\
International Business & 0.0412 & 0.1426** & -0.2166*** & -0.1041 \\
Labor and Wages & 0.4373*** & 0.2587*** & -0.3064*** & -0.5700*** \\
Mergers and Acquisitions & 0.3394*** & 0.2887*** & -0.1444 & -0.3737*** \\
Online Reviews & 0.4344*** & 0.3183*** & -0.2309*** & -0.3616*** \\
Personnel Economics & 0.1626** & 0.2222** & -0.4832*** & -0.1993*** \\
Platform Economics & -0.1744** & -0.1286 & 0.4856*** & 0.3328*** \\
Public Governance & 0.0958** & 0.0681 & -0.1348** & -0.2034*** \\
Racial Inequality & -0.1544*** & 0.1667*** & 0.4202*** & 0.5214*** \\
Retail Pricing & 0.2858*** & 0.2333** & -0.2660** & -0.5658*** \\
Risk Preferences & 0.2476*** & 0.1725* & -0.2647*** & -0.3558*** \\
Supply Chain Economics & -0.2574** & -0.2592** & 0.0841 & 0.5233*** \\
Team Productivity & 0.2266*** & 0.1515*** & -0.3101*** & -0.4067*** \\
Venture Capital & -0.1531 & -0.1752*** & -0.0927 & -0.2074* \\
\midrule
Forum Indicators & Yes & Yes & Yes & Yes \\
Year Fixed Effects & Yes & Yes & Yes & Yes \\
Thread-Title Emotion Controls & Yes & Yes & Yes & Yes \\
Observations & 1,255,942 & 1,255,942 & 1,255,942 & 1,255,942 \\
Pseudo $R^2$ & 0.0022 & 0.0018 & 0.0043 & 0.0043 \\
\bottomrule
\end{longtable}
\endgroup

\noindent\footnotesize Notes: dependent variables are reply-level sentiment scores on the $[0,1]$ scale. The first two columns measure openness and curiosity, for which higher values indicate greater openness and curiosity; the last two columns measure negative tone and poisonousness, for which higher values indicate more negative and poisonous language. Standard errors are two-way clustered by calendar year and thread. * $p<0.1$, ** $p<0.05$, *** $p<0.01$.
\normalsize

\subsection{AI-Topicality Openness Dynamics Over Time}

We next use the structured AI-topicality and attitude measures in year-specific interactions. Let $D_{t,y}$ indicate calendar year $y$; the output reports the AI-topicality interaction for each available year:

\begin{equation}
Y_{i,t} = \beta_0 + \sum_{y \ne 2012} \alpha_y \cdot \left(\mathrm{Topicality}_{AI,i} \times D_{t,y}\right) + \sum_k \beta_k \cdot \mathrm{Topicality}_{i,k} + \sum_{y \ne 2012} \gamma_y D_{t,y} + X_i' \delta + \epsilon_{i,t}.
\label{eq:dynamic_year}
\end{equation}

The $\gamma_y$ coefficients absorb calendar-year fixed effects, and each $\alpha_y$ measures the deviation of the AI-topicality gradient in year $y$ from the 2012 baseline, conditional on the included controls. These are descriptive time estimates.

Relative to the 2012 AI-topicality baseline, the estimates trace a descriptive U-shaped pattern: AI-related discussion becomes less open and less curious, and more negative and poisonous, year by year through the earlier period. The pattern then improves. Openness begins to recover in 2018 and is relatively more favorable from 2022 onward; curiosity begins to recover in 2021 and rises further in 2023--2024. Negative tone and poisonousness both begin to decline in 2018 and continue to improve in the later years. Taken together, the results indicate that AI-related EJMR discussion has become more favorable in recent years. This pattern is consistent with a technology-adoption interpretation of how AI is discussed on EJMR, but the estimates are descriptive associations rather than causal evidence of adoption.

\begin{figure}[H]
	\centering
	\includegraphics[width=0.96\textwidth]{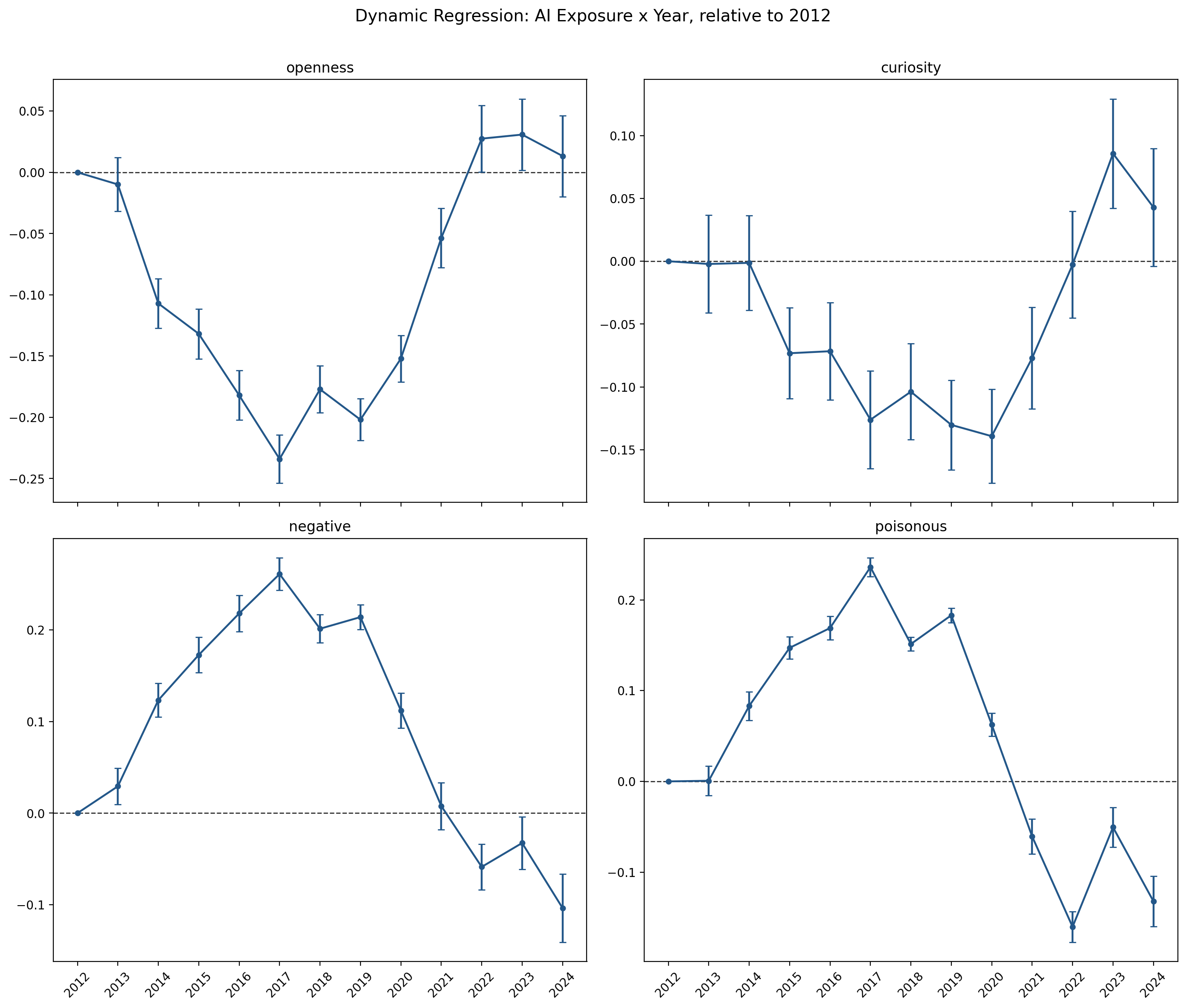}
\caption{Dynamic AI-Topicality Coefficients Relative to 2012}
	\label{fig:dynamic_ai}
\end{figure}

\normalsize

\subsection{Comparative Topic Results: Platform Economics and Gender Economics}
\label{sec:comparative_topic}

Finally, the comparative figures report year interactions for the structured topicality measures for Platform Economics and Gender Economics. Their distinct trajectories show that the results capture topic-specific patterns in the EJMR archive rather than simply a common time trend. Platform Economics resembles the earlier AI pattern in 2020, with lower openness and higher negative tone and poisonousness, but these associations largely fade by 2024 and negative tone reverses sign. Gender Economics remains associated with lower openness across years, while its poisonousness association increases markedly in 2023. These comparisons help situate the AI U-shape as a topic-specific descriptive pattern; they do not establish a general adoption mechanism.

\begin{figure}[H]
    \centering
    \begin{minipage}[t]{0.48\textwidth}
        \centering
        \includegraphics[width=\linewidth]{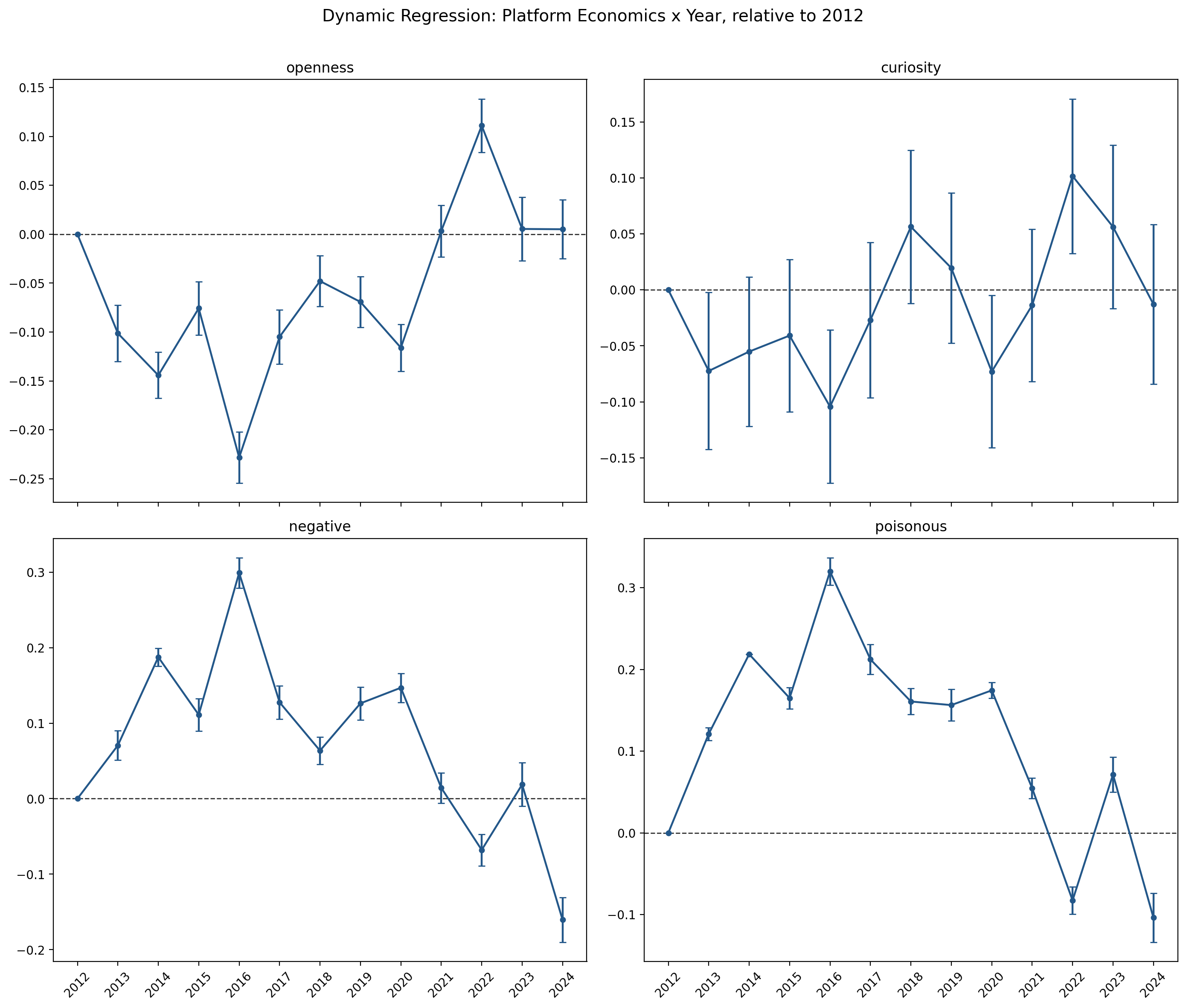}
        \captionof{figure}{Platform Economics}
        \label{fig:dynamic_platform}
    \end{minipage}\hfill
    \begin{minipage}[t]{0.48\textwidth}
        \centering
        \includegraphics[width=\linewidth]{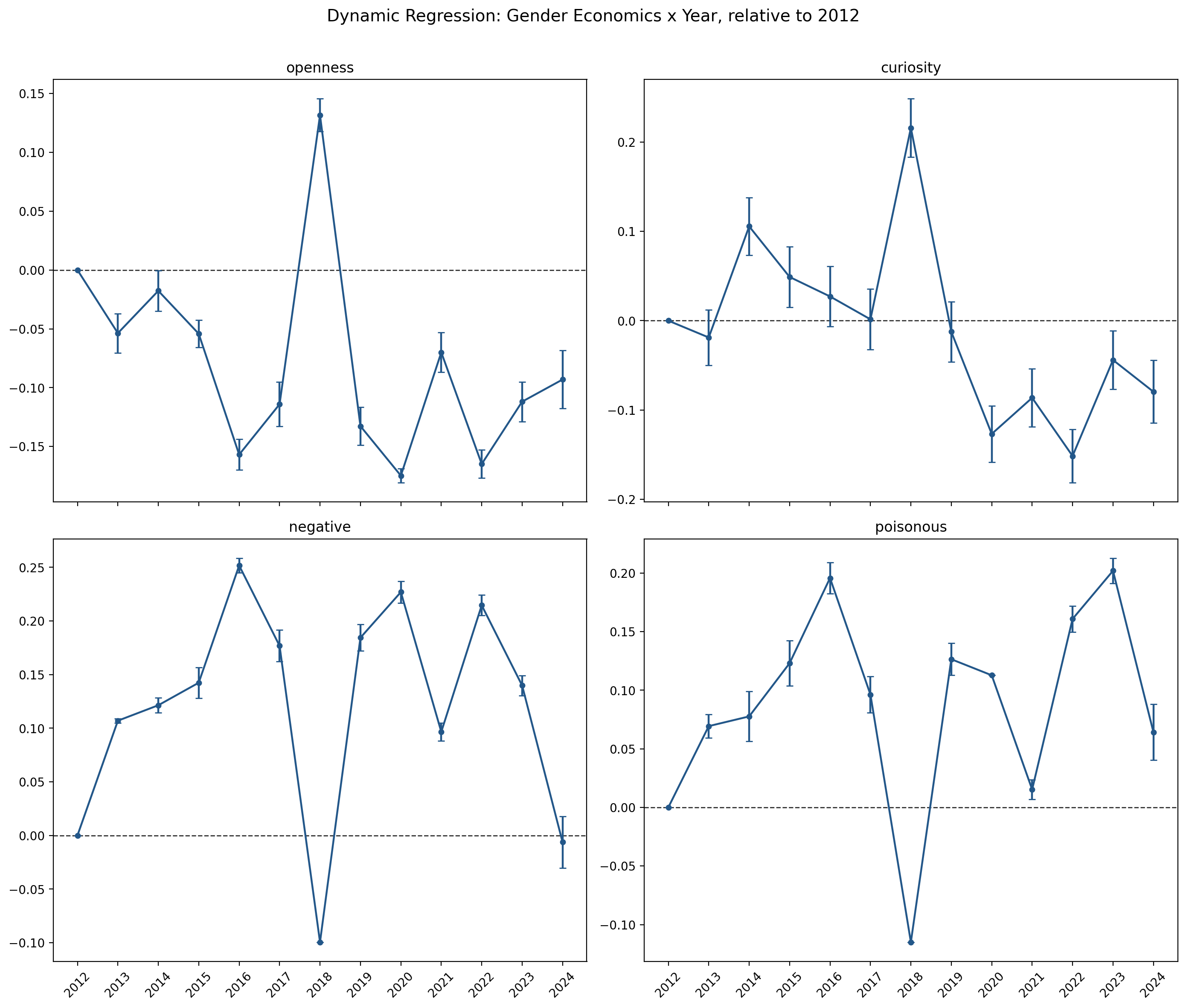}
        \captionof{figure}{Gender Economics}
        \label{fig:dynamic_gender}
    \end{minipage}
\end{figure}

\section{Conclusions}
\label{sec:conclusion}
Text as Data as Survey (TaDaS) is a method contribution for recovering survey-like, reference-anchored measures from naturally occurring text. Its core discipline is to distinguish the anchor corpus that defines interpretable semantic coordinates from the response corpus that supplies observations, and to keep outcome scoring separate from measurement. This architecture provides a transparent basis for studying technological, organizational, and market change when direct surveys are costly, unavailable retrospectively, or vulnerable to self-presentation bias.

The EJMR--AI demonstration illustrates what that architecture can reveal without overstating its reach. Elite-journal publications serve as the anchor corpus and EJMR replies as the response corpus. Within this platform-specific, nonrepresentative archive, AI-related discussion is less open and more negative than other research-related discussion in the static specification. The year-interaction estimates show that this early pattern initially worsens but reverses in recent years: openness begins to recover in 2018, curiosity in 2021, and negative tone and poisonousness begin to decline in 2018. Thus, the measured AI-related discourse becomes more favorable overall in the later period. This evolution is consistent with a technology-adoption interpretation, but it is not causal evidence that adoption generated the change.

The same design can transfer to organizational and market settings in which a structured corpus can anchor naturally occurring responses: investor discussions, customer reviews, employee feedback, public comments, and education all offer potential response corpora. Each context requires validation of its anchors, scores, and substantive interpretation, and TaDaS does not make any archive representative. Used with those safeguards, the framework enables disciplined comparison of how stakeholders discuss and evaluate change across settings.
\clearpage
\appendix
\section*{Appendix}
\renewcommand{\thetable}{A\arabic{table}}
\renewcommand{\theHtable}{A\arabic{table}}
\setcounter{table}{0}

\section{Illustrative Topic Clusters}
\label{app:topic_clusters}

This appendix reports illustrative anchor-corpus examples for the learned topic clusters. The construction of topic keywords, anchor-corpus topic centers, and response-corpus topicality measures is documented in Online Appendix.

\begingroup
\scriptsize
\setlength{\LTcapwidth}{\textwidth}
\begin{longtable}{p{0.18\textwidth}p{0.72\textwidth}}
\caption{Illustrative Publication Cases by Topic}\label{tab:paper_topic_cases}\\
\toprule
Topic & Illustrative paper titles \\
\midrule
\endfirsthead
\toprule
Topic & Illustrative paper titles \\
\midrule
\endhead
AI & \casepair{Iterative Alternative Evaluation Within Human-Artificial Intelligence Problem Solving: An Extension to Raisch and Fomina's ``Combining Human and Artificial Intelligence''}{Algorithmic Interactions in Open Source Work} \\
Econometrics & \casepair{Post-Selection and Post-Regularization Inference in Linear Models with Many Controls and Instruments}{Bootstrap Inference in Partially Identified Models Defined by Moment Inequalities: Coverage of the Identified Set} \\
Health Economics & \casepair{Are Drugs Substitutes or Complements for Intensive (and Expensive) Medical Treatment}{Will Public Sector Retiree Health Benefit Plans Survive? Economic and Policy Implications of Unfunded Liabilities} \\
Family Business & \casepair{Socioemotional Wealth and Proactive Stakeholder Engagement: Why Family-Controlled Firms Care More About Their Stakeholders}{When More Is Better: Multifamily Firms and Firm Performance} \\
Labor and Wages & \casepair{A firm's external environment and the hiring of a non-standard workforce: implications for organisations}{Offshoring, Transition, and Training: Evidence from Danish Matched Worker-Firm Data} \\
Gender Economics & \casepair{Sexual Harassment at Work: A Decade (Plus) of Progress}{Thinking Like a Feminist: What Feminist Theory Has to Offer Sociology} \\
Personnel Economics & \casepair{For the Want of a Nail: The Interaction of Managerial Capacity and Human Resource Management on Organizational Performance}{REFLECTIONS ON THE 2014 DECADE AWARD: IS THERE STRENGTH IN THE CONSTRUCT OF HR SYSTEM STRENGTH?} \\
Public Governance & \casepair{``We're not there to lead'': Professional roles and responsibilities in ``citizen-led'' co-production}{Putting Conflict Where It Belongs: A Response to ``Creating Shared Responsibility through Respect for Military Culture: The Russian and American Cases''} \\
Firm Strategy & \casepair{Strategic Management and Performance in Public Organizations: Findings from the Miles and Snow Framework}{A Proposal to Organize and Promote Replications} \\
Behavioral Economics & \casepair{Mutualistic Coupling Between Vocabulary and Reasoning Supports Cognitive Development During Late Adolescence and Early Adulthood}{Hunger as a Context: Food Seeking That Is Inhibited During Hunger Can Renew in the Context of Satiety} \\
Venture Capital & \casepair{(Non-)Precautionary Cash Hoarding and the Evolution of Growth Firms}{A Rift in the Ground: Theorizing the Evolution of Anchor Values in Crowdfunding Communities through the Oculus Rift Case Study} \\
Entrepreneurship & \casepair{Legitimately distinct entrepreneurial stories in evolving market categories}{Liquidity, Risk, and Occupational Choices} \\
Supply Chain Economics & \casepair{Cascades and Fluctuations in an Economy with an Endogenous Production Network}{Complementarity, Capabilities, and the Boundaries of the Firm: The Impact of Within-Firm and Interfirm Expertise on Concurrent Sourcing of Complementary Components} \\
Game Theory & \casepair{Carrot or Stick? The Evolution of Reciprocal Preferences in a Haystack Model}{The Law of the Few} \\
Mergers and Acquisitions & \casepair{Advancing Firm Growth Research: A Focus on Growth Mode Instead of Growth Rate}{How do US firms grow? New evidence from a growth decomposition} \\
Executive Compensation & \casepair{CEO Compensation and Board Structure Revisited}{That Could Have Been Me: Director Deaths, CEO Mortality Salience, and Corporate Prosocial Behavior} \\
Online Reviews & \casepair{Is More Structure Really Better? A Comparison of Frame-of-Reference Training and Descriptively Anchored Rating Scales to Improve Interviewers' Rating Quality}{Evaluating Frame-of-Reference Rater Training Effectiveness Using Performance Schema Accuracy} \\
Platform Economics & \casepair{Welcome contributor or no price competitor? The competitive interaction of free and priced technologies}{Evolutionary Competition in Platform Ecosystems} \\
Innovation and Patents & \casepair{Historical Disease Prevalence, Cultural Values, and Global Innovation}{Green innovation and firms' financial and environmental performance: The roles of pollution prevention versus control} \\
International Business & \casepair{Taking root in fertile ground: Community context and the agglomeration of hybrid companies}{Location of Decision Rights Within Multinational Firms} \\
Credit Risk & \casepair{Is Age a Determinant of SMEs' Financing Decisions? Empirical Evidence Using Panel Data Models}{Franchisees and Loan Default on Third-Party Guarantee Loans: Evidence From the United States} \\
Business Ethics & \casepair{Stretching the Moral Gray Zone: Positive Affect, Moral Disengagement, and Dishonesty}{Etiquette and Effort: Holding Doors for Others} \\
Team Productivity & \casepair{Top management team role structure: A vantage point for advancing upper echelons research}{Reframing talent identification as a status-organising process: Examining talent hierarchies through data mining} \\
Advertising Economics & \casepair{Spatiotemporal Allocation of Advertising Budgets}{Advertising Effects in Presidential Elections} \\
Retail Pricing & \casepair{Demand agglomeration economies, neighbor heterogeneity, and firm survival: The effect of HHGregg's bankruptcy}{Demand agglomeration economies, neighbor heterogeneity, and firm survival: The effect of HHGregg's bankruptcy} \\
Racial Inequality & \casepair{Unraveling Complexities of Latino Racialization}{The Long-Run Effect of Mexican Immigration on Crime in US Cities: Evidence from Variation in Mexican Fertility Rates} \\
Intergenerational Mobility & \casepair{Culture and Durable Inequality}{Cash or Condition? Evidence from a Cash Transfer Experiment} \\
Risk Preferences & \casepair{Measuring Probabilistic Risk Attitudes}{Human Decision Making in Dynamic Resource Allocation} \\
Consumer Demand & \casepair{The Joint Sales Impact of Frequency Reward and Customer Tier Components of Loyalty Programs}{Lead Offer Spillovers} \\
Brand Marketing & \casepair{Polishing the Gilt Edge: Elite Category Endurance and Symbolic Boundaries in US Luxury Hotels, 1790--2015}{Consumer Reactions to Brand Extensions in a Competitive Context: Does Fit Still Matter?} \\
Hedge Funds & \casepair{The term structure and inflation uncertainty}{Monetary Policy and Rational Asset Price Bubbles} \\
Earnings Management & \casepair{Disclosure Substitution}{Tacit Collusion and Voluntary Disclosure: Theory and Evidence from the US Automotive Industry} \\
\bottomrule
\end{longtable}
\endgroup

\section{Sentiment Scoring}
\subsection{Scoring Prompt}
\label{app:sentiment_prompt}

The decoder is instructed to score only the reply, to use the initiating post only as context, to anchor each dimension on a common 0--1 scale, and to return a fixed JSON schema. The full prompt template is reproduced below.

\begin{Verbatim}[fontsize=\footnotesize]
Analyze the economics dialogue and return JSON.

Task:
1. Evaluate ONLY A's attitude and affect.
2. Use Q only as contextual reference for interpreting A (do NOT score Q itself).
3. Use the bidirectional 0-1 scale below and keep explanation consistent with scores.

Global scale meaning (for every dimension):
- 0.0 = strong opposite pole
- 0.5 = no clear signal / neutral for this dimension
- 1.0 = strong presence of this dimension

Dimension anchors:
- openness:
  0.0 = strongly closed / defensive / conservative stance
  0.5 = neutral or unclear openness
  1.0 = strongly open / receptive / willing to engage
- negative:
  0.0 = clearly non-negative / positive tone
  0.5 = neither clearly positive nor negative
  1.0 = strongly negative tone
- poisonous:
  0.0 = clearly harmless / supportive
  0.5 = neither clearly toxic nor supportive
  1.0 = strongly toxic / hostile / harmful
- curiosity:
  0.0 = clearly disinterested / rejecting exploration
  0.5 = no clear curiosity signal
  1.0 = strongly curious / exploratory

Consistency rules:
- 0.5 means "no signal", not "medium intensity".
- If A is empty or missing, return 0.5 for all four scores.
- Do not raise poisonous/negative just because Q is toxic.
- Explanation should mention how A responds to Q as context.
- Keep explanation concise.

Provided Top Topic:
{TOP_TOPIC}

Return strictly JSON with exactly these keys:
{
  "openness": 0.5,
  "negative": 0.5,
  "poisonous": 0.5,
  "curiosity": 0.5,
  "explanation": ""
}

Q: {Q}
A: {A}
\end{Verbatim}

\subsection{Scoring Cases}
\begin{table}[H]
    \centering
    \caption{Illustrative Reply-Level Scoring Checks}
    \label{tab:scoring_audit_cases}
    \scriptsize
    \setlength{\tabcolsep}{3pt}
    \renewcommand{\arraystretch}{1.08}
    \begin{tabularx}{\textwidth}{p{0.24\textwidth} X p{0.19\textwidth} p{0.24\textwidth}}
        \toprule
        Thread Title & Reply Excerpt & Scores & Reasoning from LLM \\
        \midrule
        How useful is Bayesian econometrics? &
        ``To be frank, do not worship Bayesian methods, but instead treat them as a benchmark tractable learning scheme. We can then modify them, for example by incorporating recency bias; this would be quite interesting.'' &
        Openness 0.8; curiosity 0.7; negative 0.2; poisonousness 0.1 &
        The reply proposes modifications and calls the direction interesting, indicating openness and curiosity, while the ``do not worship'' phrasing contributes mild negativity and poisonousness. \\
        \addlinespace
        Why is probability theory/stochastic calculus out of fashion in the finance world? &
        ``Why? Do funds not really use derivatives anymore?'' &
        Openness 0.7; curiosity 0.8; negative 0.3; poisonousness 0.2 &
        The reply engages with the question through a clarifying follow-up, showing curiosity and openness; its challenge to the premise creates modest negative and poisonous tone. \\
        \addlinespace
        Anyone tried using models to manage his own portfolio? &
        ``It is not your model, but the quality of inputs of your model matters. And getting quality inputs is a professional effort, not a hobby. Just buy an index, please. And use your smartness to create a real career.'' &
        Openness 0.3; curiosity 0.2; negative 0.7; poisonousness 0.6 &
        The reply dismisses the inquiry as unsuitable for a hobbyist and recommends passive investing, yielding low openness and curiosity together with substantial negative and poisonous tone. \\
        \addlinespace
        Causality vs Machine Learning &
        ``You clearly don't understand not only economics or econometrics, but basic first-year statistics.'' &
        Openness 0.0; curiosity 0.0; negative 1.0; poisonousness 1.0 &
        The reply is directly dismissive and condescending, producing the lowest openness and curiosity scores and the highest negative and poisonousness scores. \\
        \bottomrule
    \end{tabularx}
\end{table}
\section{AI Usage Statement}
\label{app:ai_usage_statement}

The authors used AI-based writing tools to assist with language polishing, including improving clarity, grammar, and presentation. The research question, empirical design, data construction, regression results, statistical analysis, interpretation of results, and all substantive conclusions are the authors' own. The authors reviewed and are responsible for the final content of the paper.

\clearpage
\bibliographystyle{plainnat}
\bibliography{0_OpenEcon_refs}

@article{acemoglu2022,
  author = {Acemoglu, Daron and Autor, David and Hazell, Jonathon and Restrepo, Pascual},
  year = {2022},
  title = {Artificial Intelligence and Jobs: Evidence from Online Vacancies},
  journal = {Journal of Labor Economics},
  volume = {40},
  number = {S1},
  pages = {S293--S340},
  doi = {10.1086/718327},
  note = {NBER Working Paper No. 28257}
}

@article{alberti2024,
  author = {Alberti, Cristina T. and Morris, L.},
  year = {2024},
  title = {Online Toxic Communication about the Accounting Academic Job Market},
  journal = {Accounting Horizons},
  volume = {38},
  number = {1},
  pages = {7--26},
  doi = {10.2308/horizons-2022-066}
}

@article{autor2015,
  author = {Autor, David H.},
  year = {2015},
  title = {Why Are There Still So Many Jobs? The History and Future of Workplace Automation},
  journal = {Journal of Economic Perspectives},
  volume = {29},
  number = {3},
  pages = {3--30},
  doi = {10.1257/jep.29.3.3}
}

@article{baker2016,
  author = {Baker, Scott R. and Bloom, Nicholas and Davis, Steven J.},
  year = {2016},
  title = {Measuring Economic Policy Uncertainty},
  journal = {Quarterly Journal of Economics},
  volume = {131},
  number = {4},
  pages = {1593--1636},
  doi = {10.1093/qje/qjw024}
}

@article{bello2025,
  author = {Bello, Paolo and Casarico, Alessandra and Nozza, Debora},
  year = {2025},
  title = {Research Similarity and Women in Academia},
  journal = {Economic Journal},
  volume = {ueaf113},
  doi = {10.1093/ej/ueaf113}
}

@article{bertrand2001,
  author = {Bertrand, Marianne and Mullainathan, Sendhil},
  year = {2001},
  title = {Do People Mean What They Say? Implications for Subjective Survey Data},
  journal = {American Economic Review},
  volume = {91},
  number = {2},
  pages = {67--72},
  doi = {10.1257/aer.91.2.67}
}

@article{blei2003,
  author = {Blei, David M. and Ng, Andrew Y. and Jordan, Michael I.},
  year = {2003},
  title = {Latent Dirichlet Allocation},
  journal = {Journal of Machine Learning Research},
  volume = {3},
  pages = {993--1022},
  url = {https://www.jmlr.org/papers/v3/blei03a.html}
}

@article{bourdieu1975,
  author = {Bourdieu, Pierre},
  year = {1975},
  title = {The Specificity of the Scientific Field and the Social Conditions of the Progress of Reason},
  journal = {Social Science Information},
  volume = {14},
  number = {6},
  pages = {19--47},
  doi = {10.1177/053901847501400602}
}

@inproceedings{brown2020,
  author = {Brown, Tom B. and Mann, Benjamin and Ryder, Nick and Subbiah, Melanie and Kaplan, Jared and Dhariwal, Prafulla and Neelakantan, Arvind and Shyam, Pranav and Sastry, Girish and Askell, Amanda and Agarwal, Sandhini and Herbert-Voss, Ariel and Krueger, Gretchen and Henighan, Tom and Child, Rewon and Ramesh, Aditya and Ziegler, Daniel M. and Wu, Jeffrey and Winter, Clemens and Hesse, Christopher and Chen, Mark and Sigler, Eric and Litwin, Mateusz and Gray, Scott and Chess, Benjamin and Clark, Jack and Berner, Christopher and McCandlish, Sam and Radford, Alec and Sutskever, Ilya and Amodei, Dario},
  year = {2020},
  title = {Language Models are Few-Shot Learners},
  booktitle = {Advances in Neural Information Processing Systems},
  volume = {33},
  pages = {1877--1901},
  note = {arXiv:2005.14165}
}

@article{brynjolfsson2025,
  author = {Brynjolfsson, Erik and Li, Danielle and Raymond, Lindsey},
  year = {2025},
  title = {Generative {AI} at Work},
  journal = {Quarterly Journal of Economics},
  volume = {140},
  number = {2},
  pages = {889--942},
  doi = {10.1093/qje/qjae044}
}

@inproceedings{campello2013hdbscan,
  author = {Campello, Ricardo J. G. B. and Moulavi, Davoud and Sander, Joerg},
  year = {2013},
  title = {Density-Based Clustering Based on Hierarchical Density Estimates},
  booktitle = {Advances in Knowledge Discovery and Data Mining},
  series = {Lecture Notes in Computer Science},
  volume = {7819},
  pages = {160--172},
  doi = {10.1007/978-3-642-37456-2_14}
}

@inproceedings{cheng2017,
  author = {Cheng, Justin and Bernstein, Michael and Danescu-Niculescu-Mizil, Cristian and Leskovec, Jure},
  year = {2017},
  title = {Anyone Can Become a Troll: Causes of Trolling Behavior in Online Discussions},
  booktitle = {Proceedings of the 2017 ACM Conference on Computer Supported Cooperative Work and Social Computing (CSCW)},
  pages = {1217--1230},
  doi = {10.1145/2998181.2998213}
}

@book{crane1972,
  author = {Crane, Diana},
  year = {1972},
  title = {Invisible Colleges: Diffusion of Knowledge in Scientific Communities},
  publisher = {University of Chicago Press},
  address = {Chicago},
  isbn = {978-0226118576}
}

@inproceedings{devlin2019bert,
  author = {Devlin, Jacob and Chang, Ming-Wei and Lee, Kenton and Toutanova, Kristina},
  year = {2019},
  title = {{BERT}: Pre-training of Deep Bidirectional Transformers for Language Understanding},
  booktitle = {Proceedings of the 2019 Conference of the North American Chapter of the Association for Computational Linguistics: Human Language Technologies (NAACL-HLT)},
  pages = {4171--4186},
  note = {arXiv:1810.04805}
}

@article{ederer2025,
  author = {Ederer, Florian and Goldsmith-Pinkham, Paul and Jensen, Kyle},
  year = {2025},
  title = {Anonymous Attention and Abuse},
  journal = {AEA Papers and Proceedings},
  volume = {115},
  pages = {188--194},
  doi = {10.1257/pandp.20251048}
}

@misc{eloundou2023,
  author = {Eloundou, Tyna and Manning, Sam and Mishkin, Pamela and Rock, Daniel},
  year = {2023},
  title = {{GPTs} are {GPTs}: An Early Look at the Labor Market Impact Potential of Large Language Models},
  note = {arXiv:2303.10130}
}

@misc{fan2024,
  author = {Fan, Jiahua and Liu, Qiang and Song, Yifei and Wang, Zhuo},
  year = {2024},
  title = {Measuring Misinformation in Financial Markets},
  note = {SSRN Working Paper No. 4922648},
  doi = {10.2139/ssrn.4922648}
}

@article{felten2021,
  author = {Felten, Edward W. and Raj, Manav and Seamans, Robert},
  year = {2021},
  title = {Occupational, Industry, and Geographic Exposure to Artificial Intelligence: A Novel Dataset and Its Potential Uses},
  journal = {Strategic Management Journal},
  volume = {42},
  number = {12},
  pages = {2195--2217},
  doi = {10.1002/smj.3286}
}

@article{flores2023,
  author = {Flores, Pamela M. and Hilbert, Martin},
  year = {2023},
  title = {Lean-back and Lean-forward Online Behaviors: The Role of Emotions in Passive versus Proactive Information Diffusion of Social Media Content},
  journal = {Computers in Human Behavior},
  volume = {148},
  pages = {107841},
  doi = {10.1016/j.chb.2023.107841}
}

@article{gentzkow2019,
  author = {Gentzkow, Matthew and Kelly, Bryan and Taddy, Matt},
  year = {2019},
  title = {Text as Data},
  journal = {Journal of Economic Literature},
  volume = {57},
  number = {3},
  pages = {535--574},
  doi = {10.1257/jel.20181020}
}

@article{gilardi2023,
  author = {Gilardi, Fabrizio and Alizadeh, Meysam and Kubli, Mael},
  year = {2023},
  title = {{ChatGPT} Outperforms Crowd Workers for Text-Annotation Tasks},
  journal = {Proceedings of the National Academy of Sciences},
  volume = {120},
  number = {30},
  pages = {e2305016120},
  doi = {10.1073/pnas.2305016120}
}

@misc{goldsmithpinkham2024,
  author = {Goldsmith-Pinkham, Paul},
  year = {2026},
  title = {Tracking the Credibility Revolution Across Fields},
  note = {NBER Working Paper No. 35051},
  doi = {10.3386/w35051}
}

@article{grimmer2013,
  author = {Grimmer, Justin and Stewart, Brandon M.},
  year = {2013},
  title = {Text as Data: The Promise and Pitfalls of Automatic Content Analysis Methods for Political Texts},
  journal = {Political Analysis},
  volume = {21},
  number = {3},
  pages = {267--297},
  doi = {10.1093/pan/mps028}
}

@book{groves2009,
  author = {Groves, Robert M. and Fowler, Floyd J. and Couper, Mick P. and Lepkowski, James M. and Singer, Eleanor and Tourangeau, Roger},
  year = {2009},
  title = {Survey Methodology},
  edition = {2},
  publisher = {Wiley},
  address = {Hoboken, NJ},
  isbn = {978-0-470-46546-2}
}

@misc{grootendorst2022bertopic,
  author = {Grootendorst, Maarten},
  year = {2022},
  title = {{BERTopic}: Neural Topic Modeling with a Class-Based {TF-IDF} Procedure},
  note = {arXiv:2203.05794}
}

@techreport{gemmateam2025,
  author = {{Gemma Team, Google DeepMind}},
  year = {2025},
  title = {Gemma 3 Technical Report},
  institution = {Google DeepMind},
  note = {arXiv:2503.19786}
}

@article{hao2025,
  author = {Hao, Q. and Xu, F. and Li, Y. and Evans, J.},
  year = {2026},
  title = {Artificial Intelligence Tools Expand Scientists' Impact but Contract Science's Focus},
  journal = {Nature},
  volume = {649},
  pages = {1237--1243},
  doi = {10.1038/s41586-025-09922-y}
}

@article{hassan2025,
  author = {Hassan, Tarek A. and Hollander, Stephan and Kalyani, A. and van Lent, Laurence and Schwedeler, M. and Tahoun, Ahmed},
  year = {2025},
  title = {Text as Data in Economic Analysis},
  journal = {Journal of Economic Perspectives},
  volume = {39},
  number = {3},
  pages = {193--220},
  doi = {10.1257/jep.20231365}
}

@article{hoberg2016,
  author = {Hoberg, Gerard and Phillips, Gordon},
  year = {2016},
  title = {Text-Based Network Industries and Endogenous Product Differentiation},
  journal = {Journal of Political Economy},
  volume = {124},
  number = {5},
  pages = {1423--1465},
  doi = {10.1086/688176}
}

@techreport{kanazawa2025,
  author = {Kanazawa, Kyogo and Kawaguchi, Daiji and Shigeoka, Hitoshi and Watanabe, Y.},
  year = {2025},
  title = {{AI}, Skill, and Productivity: The Case of Taxi Drivers},
  institution = {National Bureau of Economic Research},
  number = {30612},
  note = {Revised 2025; SSRN:4260700},
  doi = {10.3386/w30612}
}

@article{li2024,
  author = {Chen, L. and Mankad, S.},
  year = {2025},
  title = {A Structural Topic and Sentiment-Discourse Model for Text Analysis},
  journal = {Management Science},
  volume = {71},
  number = {7},
  pages = {5767--5787},
  doi = {10.1287/mnsc.2022.00261}
}

@article{lin2024,
  author = {Cong, Lin William and Liang, T. and Zhang, X. and Zhu, W.},
  year = {2025},
  title = {Textual Factors: A Scalable, Interpretable, and Data-Driven Approach to Analyzing Unstructured Information},
  journal = {Management Science},
  volume = {71},
  number = {12},
  pages = {10727--10739},
  doi = {10.1287/mnsc.2020.01180}
}

@inproceedings{locatelli2023,
  author = {Locatelli, M. S. and Calais, P. H. and Miranda, M. P. and Junho, J. P. and Muniz, T. L. and Meira Jr., Wagner and Almeida, Virgilio},
  year = {2024},
  title = {Topic Shifts as a Proxy for Assessing Politicization in Social Media},
  booktitle = {Proceedings of the International AAAI Conference on Web and Social Media (ICWSM)},
  volume = {18},
  pages = {972--984},
  doi = {10.1609/icwsm.v18i1.31366},
  note = {arXiv:2312.11326}
}

@article{loughran2011,
  author = {Loughran, Tim and McDonald, Bill},
  year = {2011},
  title = {When Is a Liability Not a Liability? Textual Analysis, Dictionaries, and 10-Ks},
  journal = {Journal of Finance},
  volume = {66},
  number = {1},
  pages = {35--65},
  doi = {10.1111/j.1540-6261.2010.01625.x}
}

@article{mcinnes2017hdbscan,
  author = {McInnes, Leland and Healy, John and Astels, Steve},
  year = {2017},
  title = {hdbscan: Hierarchical Density Based Clustering},
  journal = {Journal of Open Source Software},
  volume = {2},
  number = {11},
  pages = {205},
  doi = {10.21105/joss.00205}
}

@article{mcinnes2018umap,
  author = {McInnes, Leland and Healy, John and Saul, Nathaniel and Grossberger, Lukas},
  year = {2018},
  title = {{UMAP}: Uniform Manifold Approximation and Projection},
  journal = {Journal of Open Source Software},
  volume = {3},
  number = {29},
  pages = {861},
  doi = {10.21105/joss.00861}
}

@article{merton1968,
  author = {Merton, Robert K.},
  year = {1968},
  title = {The Matthew Effect in Science: The Reward and Communication Systems of Science Are Considered},
  journal = {Science},
  volume = {159},
  number = {3810},
  pages = {56--63},
  doi = {10.1126/science.159.3810.56}
}

@misc{liu2019roberta,
  author = {Liu, Yinhan and Ott, Myle and Goyal, Naman and Du, Jingfei and Joshi, Mandar and Chen, Danqi and Levy, Omer and Lewis, Mike and Zettlemoyer, Luke and Stoyanov, Veselin},
  year = {2019},
  title = {{RoBERTa}: A Robustly Optimized {BERT} Pretraining Approach},
  note = {arXiv:1907.11692}
}

@inproceedings{ouyang2022,
  author = {Ouyang, Long and Wu, Jeffrey and Jiang, Xu and Almeida, Diogo and Wainwright, Carroll L. and Mishkin, Pamela and Zhang, Chong and Agarwal, Sandhini and Slama, Katarina and Ray, Alex and Schulman, John and Hilton, Jacob and Kelton, Fraser and Miller, Luke and Simens, Maddie and Askell, Amanda and Welinder, Peter and Christiano, Paul and Leike, Jan and Lowe, Ryan},
  year = {2022},
  title = {Training Language Models to Follow Instructions with Human Feedback},
  booktitle = {Advances in Neural Information Processing Systems},
  volume = {35},
  pages = {27730--27744},
  note = {arXiv:2203.02155}
}

@article{noy2023,
  author = {Noy, Shakked and Zhang, Whitney},
  year = {2023},
  title = {Experimental Evidence on the Productivity Effects of Generative Artificial Intelligence},
  journal = {Science},
  volume = {381},
  number = {6654},
  pages = {187--192},
  doi = {10.1126/science.adh2586}
}

@article{pramanick2026,
  author = {Pramanick, Aniket and Hou, Yufang and Mohammad, Saif M. and Gurevych, Iryna},
  year = {2026},
  title = {Transforming Scholarly Landscapes: The Influence of Large Language Models on Academic Fields Beyond Computer Science},
  journal = {PLOS One},
  volume = {21},
  number = {1},
  pages = {e0337127},
  doi = {10.1371/journal.pone.0337127}
}

@inproceedings{reimers2019,
  author = {Reimers, Nils and Gurevych, Iryna},
  year = {2019},
  title = {Sentence-{BERT}: Sentence Embeddings using Siamese {BERT}-Networks},
  booktitle = {Proceedings of the 2019 Conference on Empirical Methods in Natural Language Processing and the 9th International Joint Conference on Natural Language Processing (EMNLP-IJCNLP)},
  pages = {3982--3992},
  doi = {10.18653/v1/D19-1410},
  note = {arXiv:1908.10084}
}

@article{roberts2014,
  author = {Roberts, Margaret E. and Stewart, Brandon M. and Tingley, Dustin and Lucas, Christopher and Leder-Luis, Jetson and Gadarian, Shana Kushner and Albertson, Bethany and Rand, David G.},
  year = {2014},
  title = {Structural Topic Models for Open-Ended Survey Responses},
  journal = {American Journal of Political Science},
  volume = {58},
  number = {4},
  pages = {1064--1082},
  doi = {10.1111/ajps.12103}
}

@article{siano2025,
  author = {Siano, Francesco},
  year = {2025},
  title = {The News in Earnings Announcement Disclosures: Capturing Word Context Using {LLM} Methods},
  journal = {Management Science},
  volume = {71},
  number = {11},
  pages = {9831--9855},
  doi = {10.1287/mnsc.2024.05417},
  note = {SSRN:5198675}
}

@article{suler2004,
  author = {Suler, John},
  year = {2004},
  title = {The Online Disinhibition Effect},
  journal = {CyberPsychology \& Behavior},
  volume = {7},
  number = {3},
  pages = {321--326},
  doi = {10.1089/1094931041291295}
}

@inproceedings{vaswani2017,
  author = {Vaswani, Ashish and Shazeer, Noam and Parmar, Niki and Uszkoreit, Jakob and Jones, Llion and Gomez, Aidan N. and Kaiser, Lukasz and Polosukhin, Illia},
  year = {2017},
  title = {Attention Is All You Need},
  booktitle = {Advances in Neural Information Processing Systems},
  volume = {30},
  pages = {5998--6008},
  note = {arXiv:1706.03762}
}

@article{tourangeau2007,
  author = {Tourangeau, Roger and Yan, Ting},
  year = {2007},
  title = {Sensitive Questions in Surveys},
  journal = {Psychological Bulletin},
  volume = {133},
  number = {5},
  pages = {859--883},
  doi = {10.1037/0033-2909.133.5.859}
}

@article{wu2018,
  author = {Wu, Alice H.},
  year = {2018},
  title = {Gendered Language on the Economics Job Market Rumors Forum},
  journal = {AEA Papers and Proceedings},
  volume = {108},
  pages = {175--179},
  doi = {10.1257/pandp.20181101}
}

@article{xiu2024,
  author = {Bybee, Leland and Kelly, Bryan and Manela, Asaf and Xiu, Dacheng},
  year = {2024},
  title = {Business News and Business Cycles},
  journal = {Journal of Finance},
  volume = {79},
  number = {5},
  pages = {3105--3147},
  doi = {10.1111/jofi.13377},
  note = {NBER Working Paper No. 29344}
}

@article{ziems2024,
  author = {Ziems, Caleb and Held, William and Shaikh, Omar and Chen, Jiaao and Zhang, Zhehao and Yang, Diyi},
  year = {2024},
  title = {Can Large Language Models Transform Computational Social Science?},
  journal = {Computational Linguistics},
  volume = {50},
  number = {1},
  pages = {237--291},
  doi = {10.1162/coli_a_00502}
}

\clearpage
\pagenumbering{arabic}
\setcounter{figure}{0}
\renewcommand{\thefigure}{OA\arabic{figure}}
\renewcommand{\theHfigure}{OA\arabic{figure}}
\setcounter{table}{0}
\renewcommand{\thetable}{OA\arabic{table}}
\renewcommand{\theHtable}{OA\arabic{table}}
\setcounter{section}{0}
\setcounter{subsection}{0}
\renewcommand{\thesection}{OA\arabic{section}}
\renewcommand{\thesubsection}{\thesection.\arabic{subsection}}
\renewcommand{\theHsection}{OA\arabic{section}}
\renewcommand{\theHsubsection}{OA\arabic{section}.\arabic{subsection}}
\renewcommand{\thesubsubsection}{\thesubsection.\arabic{subsubsection}}
\renewcommand{\theHsubsubsection}{OA\arabic{section}.\arabic{subsection}.\arabic{subsubsection}}
\section*{Online Appendix: Are Economists Open to AI? A Text-as-Data-as-Survey Approach via Language Models}

\section{Related Methods}
\label{app:transformer_background}

This appendix explains the transformer architecture that underlies both the encoder used for topic mapping and the decoder used for reply-level attitude scoring, followed by the clustering methods that convert encoder representations into topic centers.

\subsection{Transformer}

Transformers map a token sequence into contextual representations through repeated layers of self-attention and position-wise nonlinear transformation \citep{vaswani2017}. Let a tokenized input sequence be $t_1,\dots,t_n$. After token and position embeddings are added, the initial layer representation can be written as
\begin{equation}
H^{(0)}=
\begin{bmatrix}
	e(t_1)+p_1\\
	\vdots\\
	e(t_n)+p_n
\end{bmatrix}
\in \mathbb{R}^{n \times d},
\end{equation}
where $e(t_i)$ is the token embedding, $p_i$ is the positional embedding, and $d$ is the hidden dimension.

In one attention head, the layer state $H^{(\ell-1)}$ is linearly projected into queries, keys, and values:
\begin{equation}
Q^{(\ell,h)} = H^{(\ell-1)}W_Q^{(\ell,h)}, \qquad
K^{(\ell,h)} = H^{(\ell-1)}W_K^{(\ell,h)}, \qquad
V^{(\ell,h)} = H^{(\ell-1)}W_V^{(\ell,h)},
\end{equation}
where $h$ indexes the attention head and the projection matrices have compatible dimensions. Scaled dot-product attention for that head is then
\begin{equation}
\mathrm{Attn}^{(\ell,h)}(H^{(\ell-1)})
=
\mathrm{softmax}\!\left(
\frac{Q^{(\ell,h)}(K^{(\ell,h)})^\top}{\sqrt{d_k}}
\right)V^{(\ell,h)}.
\end{equation}
If there are $M$ heads, multi-head attention concatenates the head-specific outputs and applies an output projection:
\begin{equation}
\mathrm{MHA}^{(\ell)}(H^{(\ell-1)})
=
\mathrm{Concat}\!\left(
\mathrm{Attn}^{(\ell,1)},\dots,\mathrm{Attn}^{(\ell,M)}
\right)W_O^{(\ell)}.
\end{equation}

Each transformer block then combines multi-head attention with a position-wise feed-forward network, residual connections, and layer normalization. A standard pre-normalization representation is
\begin{equation}
\tilde{H}^{(\ell)}
=
H^{(\ell-1)} + \mathrm{MHA}^{(\ell)}\!\left(\mathrm{LN}(H^{(\ell-1)})\right),
\end{equation}
\begin{equation}
H^{(\ell)}
=
\tilde{H}^{(\ell)} + \mathrm{FFN}^{(\ell)}\!\left(\mathrm{LN}(\tilde{H}^{(\ell)})\right),
\end{equation}
with
\begin{equation}
\mathrm{FFN}^{(\ell)}(u)=W_2^{(\ell)}\phi\!\left(W_1^{(\ell)}u+b_1^{(\ell)}\right)+b_2^{(\ell)}.
\end{equation}
The key economic intuition for the present paper is simple: self-attention allows each token to be represented in the context of every other token in the same sequence, so the final representation is sensitive to meaning, tone, and surrounding context rather than to isolated keywords alone.

\subsection{Encoder}

Encoder models such as BERT and RoBERTa read the full input sequence bidirectionally and produce contextual representations for every token \citep{devlin2019bert,liu2019roberta}. If the final-layer encoder output for text $i$ is $H_i^{(L)} \in \mathbb{R}^{n_i \times d}$, a sentence-level embedding can be obtained through a pooling operator,
\begin{equation}
x_i = \mathrm{Pool}\!\left(H_i^{(L)}\right) \in \mathbb{R}^{d}.
\end{equation}
In the present application, the encoder is not used for token prediction. It is used to place publication text and EJMR text into the same semantic geometry, so that cosine similarity and cluster structure become interpretable measures of topic proximity.

The model used for this step is All-RoBERTa Large v1 (\texttt{all-roberta-large-v1}), which combines a RoBERTa backbone with sentence-transformer style fine-tuning for semantic similarity tasks \citep{liu2019roberta,reimers2019}. That design is especially useful here because the paper needs one shared representation space that can compare short, informal forum language to formal academic titles and abstracts.

\subsection{Decoder}

Decoder models use the same transformer ingredients but impose a causal attention mask so that token $s$ can only attend to positions $1,\dots,s$ when predicting the next token. If $M \in \mathbb{R}^{n\times n}$ is the causal mask with
\begin{equation}
M_{ab}=
\begin{cases}
	0, & b \le a,\\
	-\infty, & b>a,
\end{cases}
\end{equation}
then masked self-attention can be written as
\begin{equation}
\mathrm{MaskedAttn}(H)
=
\mathrm{softmax}\!\left(
\frac{QK^\top}{\sqrt{d_k}} + M
\right)V.
\end{equation}
This architecture yields an autoregressive factorization of the output distribution:
\begin{equation}
p(y_{1:T}\mid c)
=
\prod_{t=1}^{T} p(y_t \mid c, y_{1:t-1}),
\end{equation}
where $c$ denotes the conditioning context. Instruction tuning then adapts such a decoder to follow structured prompts and return useful task-specific outputs.

In this paper, the decoder is not used to generate free-form essays. It is used as a controlled scoring device. Given a prompt post $Q_j$ and a reply $A_j$, the model returns a vector of bounded attitude scores,
\begin{equation}
r_j = f_{\theta}(Q_j,A_j)
=
\big(r_j^{\mathrm{open}}, r_j^{\mathrm{neg}}, r_j^{\mathrm{tox}}, r_j^{\mathrm{cur}}\big),
\end{equation}
with each component constrained to $[0,1]$ by the prompting template. The decoder used for this stage is Gemma 3 12B IT (\texttt{gemma-3-12b-it}), an instruction-tuned decoder model in the Gemma family \citep{gemmateam2025}. Relative to an encoder, its advantage here is that it can condition directly on the pair $(Q_j,A_j)$ and follow explicit scoring instructions about whose attitude is being evaluated and how the scale should be interpreted.

\subsection{UMAP}
\label{app:topic_pipeline}

Let $x_i \in \mathbb{R}^{d}$ denote the encoder representation of publication text $i$. These vectors live in a high-dimensional semantic space. Direct clustering in that space is possible, but in practice it is often useful to first preserve the local neighborhood structure in a lower-dimensional representation. The pipeline therefore applies UMAP \citep{mcinnes2018umap} to obtain reduced vectors
\begin{equation}
\tilde{x}_i = U(x_i) \in \mathbb{R}^{r},
\end{equation}
where $U(\cdot)$ denotes the fitted UMAP transformation and $r \ll d$.

The basic role of UMAP is geometric rather than supervised. It attempts to preserve nearby semantic relationships while compressing the representation to a space that is easier to cluster. 

More concretely, UMAP begins by constructing a local neighborhood graph in the original embedding space. Around each publication vector $x_i$, it identifies nearby observations and converts the corresponding distances into fuzzy neighborhood weights. Let $p_{i\ell}$ denote the strength of the neighborhood relation between publication texts $i$ and $\ell$ in the original encoder space. The collection of these weights defines a fuzzy graph that approximates the local manifold structure of the anchor corpus.

In a standard formulation, one first defines a local connectivity offset $\rho_i$ and scale parameter $\sigma_i$ for each point, and then writes the directed neighborhood weight from $i$ to $\ell$ as
\begin{equation}
p_{\ell \mid i}
=
\exp\!\left(
-\frac{\max\{0,\|x_i-x_\ell\|-\rho_i\}}{\sigma_i}
\right).
\end{equation}
The symmetrized fuzzy weight can then be written as
\begin{equation}
p_{i\ell}
=
p_{\ell \mid i}+p_{i \mid \ell}-p_{\ell \mid i}p_{i \mid \ell}.
\end{equation}
This construction allows the neighborhood scale to adapt locally rather than imposing one global radius for the entire corpus.

UMAP then searches for reduced vectors $\tilde{x}_i$ whose own low-dimensional fuzzy neighborhood graph resembles the high-dimensional one as closely as possible. If $q_{i\ell}$ denotes the corresponding neighborhood weight in the reduced space, this fitting problem can be summarized in the form
\begin{equation}
\mathcal{L}_{\mathrm{UMAP}}
=
\sum_{i<\ell}
\left[
p_{i\ell}\log\!\left(\frac{p_{i\ell}}{q_{i\ell}}\right)
+ (1-p_{i\ell})\log\!\left(\frac{1-p_{i\ell}}{1-q_{i\ell}}\right)
\right].
\end{equation}

In the low-dimensional space, UMAP typically uses a smooth heavy-tailed kernel such as
\begin{equation}
q_{i\ell}
=
\left(1+a\|\tilde{x}_i-\tilde{x}_\ell\|^{2b}\right)^{-1},
\end{equation}
where $a>0$ and $b>0$ govern how quickly similarity decays with distance. The first term rewards keeping genuinely close publication texts near one another after reduction, while the second discourages unrelated texts from collapsing together. 

\subsection{HDBSCAN}

After dimensionality reduction, the reduced vectors $\tilde{x}_i$ are partitioned with HDBSCAN \citep{campello2013hdbscan,mcinnes2017hdbscan}. Let $g(i) \in \{1,\dots,K\}$ denote the cluster assignment for publication $i$, where $K$ is the resulting number of identified topic clusters. HDBSCAN is useful here because it is density-based and inherently hierarchical: it allows clusters to differ in size and shape, and it can leave some observations weakly assigned or effectively treated as noise rather than forcing every document into an equally compact partition. 

Its logic can be described in three steps. First, for each point $\tilde{x}_i$, define a core distance $c_i$ that captures the radius required to reach a target local neighborhood size. If $m$ denotes the minimum neighborhood size used by the procedure, one can write
\begin{equation}
c_i = d_m(\tilde{x}_i),
\end{equation}
where $d_m(\tilde{x}_i)$ is the distance from $\tilde{x}_i$ to its $m$-th nearest neighbor. Second, replace ordinary pairwise distance with mutual-reachability distance,
\begin{equation}
d_{\mathrm{mr}}(i,\ell)
=
\max\{c_i,\; c_\ell,\; \|\tilde{x}_i-\tilde{x}_\ell\|\}.
\end{equation}
This transformation makes thin bridges between sparse regions less influential and turns density structure into a more stable geometric object. Third, HDBSCAN builds a minimum spanning tree using these mutual-reachability distances, interprets that tree across all density thresholds, and condenses the resulting hierarchy into a set of stable clusters.

It is often useful to switch from distance to density level by defining
\begin{equation}
\lambda_{i\ell} = \frac{1}{d_{\mathrm{mr}}(i,\ell)}.
\end{equation}
As the density threshold $\lambda$ changes, connected components appear, split, and disappear. HDBSCAN therefore studies a full cluster tree indexed by density.

Formally, if cluster $C$ is born at density level $\lambda_{\mathrm{birth}}(C)$ and observation $i \in C$ remains attached to that cluster until level $\lambda_i^{\ast}$, a generic stability score can be written as
\begin{equation}
\mathrm{Stability}(C)
=
\sum_{i \in C}\left(\lambda_i^{\ast}-\lambda_{\mathrm{birth}}(C)\right).
\end{equation}
Clusters with larger stability are interpreted as more persistent density structures in the condensed hierarchy.

\subsection{Topic Similarity Construction}
\label{app:topic_similarity_construction}

After HDBSCAN identifies anchor-corpus clusters, the analysis converts them into interpretable topic summaries and reply-level topicality measures. This step links the cluster assignments in the publication corpus to the structured variables used in the empirical analysis.

Each cluster is summarized with representative keywords using class-based TF-IDF, following the logic used in BERTopic-style neural topic modeling \citep{grootendorst2022bertopic}. Let $\mathcal{D}_k$ denote the meta-document formed by concatenating the texts of all publications assigned to cluster $k$. For token $w$, let $f_{kw}$ denote its within-cluster frequency in $\mathcal{D}_k$, and let $n_w$ denote the number of cluster-level meta-documents in which token $w$ appears. A generic c-TF-IDF score can then be written as
\begin{equation}
\omega_{kw} = \frac{f_{kw}}{\sum_{v} f_{kv}} \cdot \log\!\left(\frac{K}{n_w}\right).
\end{equation}

High values of $\omega_{kw}$ indicate tokens that are common within cluster $k$ but relatively distinctive across clusters. The reported topic keywords in the main text are drawn from the highest-scoring terms under this cluster-level weighting scheme. The topic labels should therefore be read as semantic summaries of clusters rather than as manually imposed field names.

The pipeline then returns to the original encoder space to define each topic center. For cluster $k$, let
\begin{equation}
\tau_k = \frac{1}{|\mathcal{C}_k|} \sum_{i \in \mathcal{C}_k} x_i
\end{equation}
denote the mean publication embedding for that cluster, where $\mathcal{C}_k = \{i : g(i)=k\}$. The vector $\tau_k$ is the anchor-corpus topic center used as the benchmark for topicality.

Let $z_j \in \mathbb{R}^{d}$ denote the encoder representation of EJMR reply $j$. The topicality score between reply $j$ and anchor-corpus topic center $k$ is computed as cosine similarity:
\begin{equation}
s_{jk} = \cos(z_j,\tau_k)
= \frac{z_j^\top \tau_k}{\|z_j\| \, \|\tau_k\|}.
\end{equation}

These $s_{jk}$ scores are the structured topicality variables used in the descriptive analyses. In short, the analysis learns clusters from the anchor corpus, represents each cluster by its mean semantic location, and measures the proximity of each EJMR reply to those anchor-corpus topic centers.

\section{Supplementary Figures}

\subsection{EJMR Response-Corpus Context}

\begin{figure}[H]
	\centering
	\includegraphics[width=0.88\textwidth]{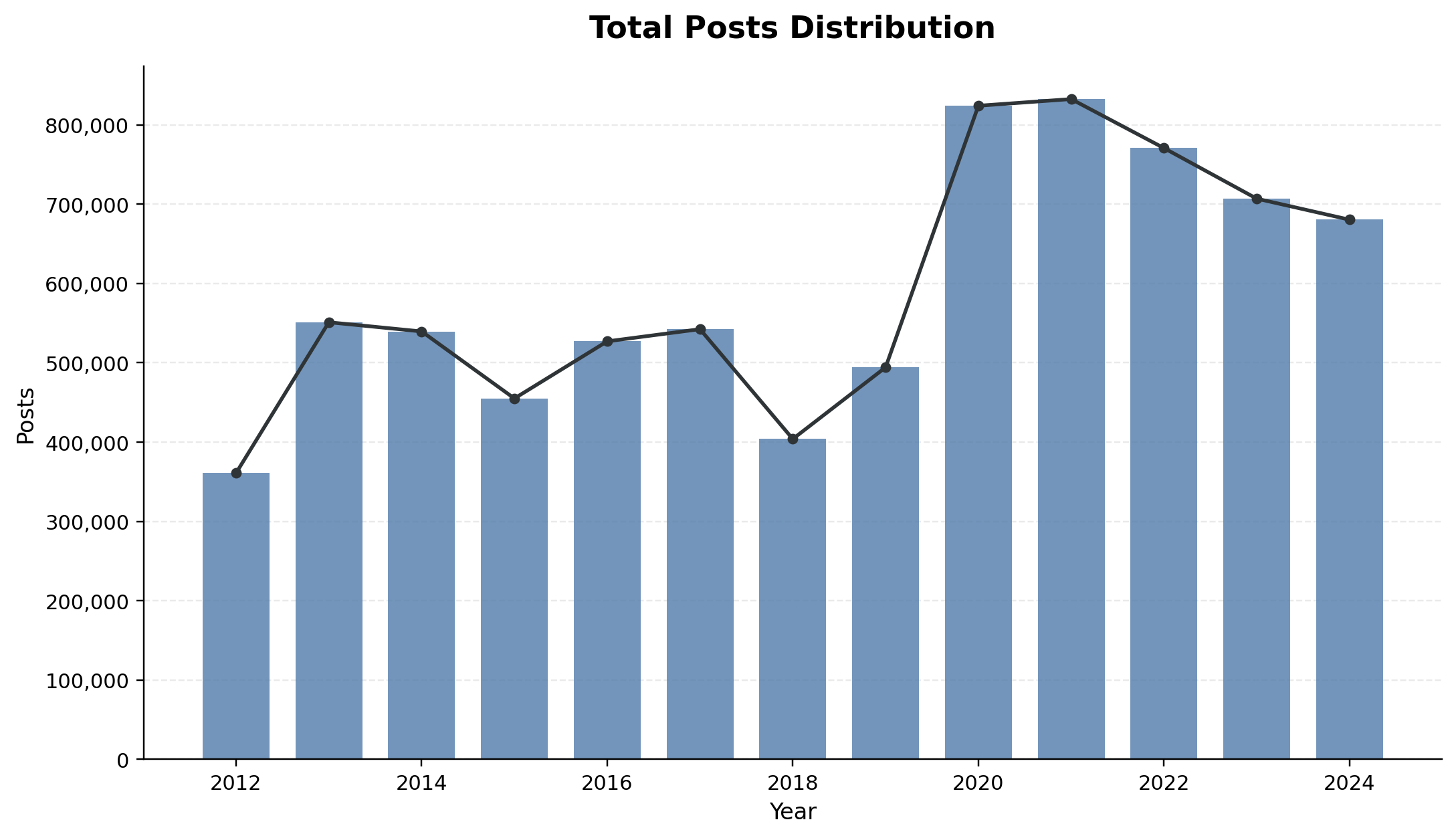}
	\caption{Full EJMR Response-Corpus Posts Over Time}
	\label{fig:ejmr_topic_distribution_full}
\end{figure}

\begin{figure}[H]
	\centering
	\includegraphics[width=0.82\textwidth]{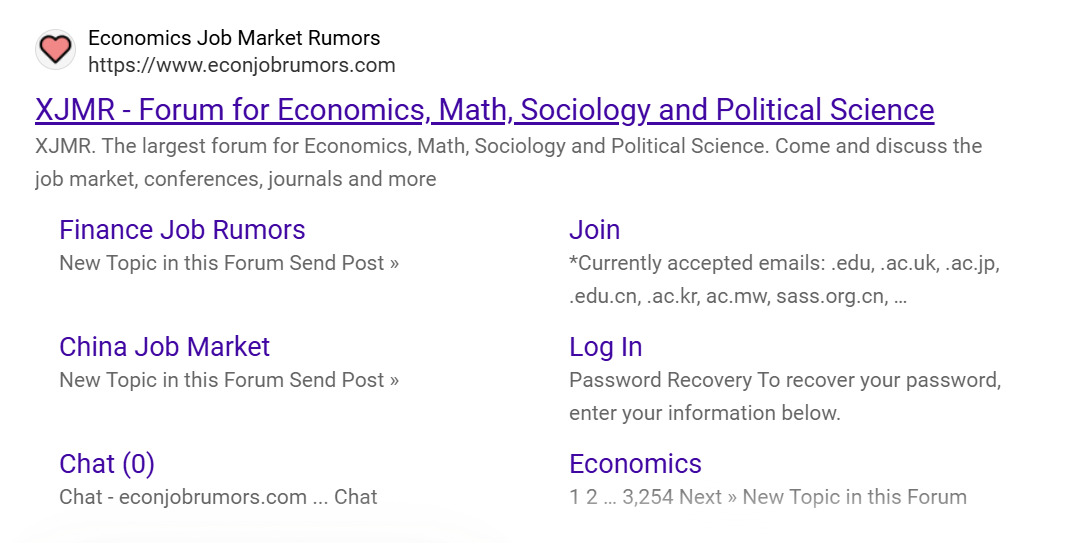}
	\caption{EJMR Response-Corpus Discussion Environment}
	\label{fig:ejmr_frontpage}
\end{figure}

\begin{figure}[H]
	\centering
	\begin{minipage}[t]{0.31\textwidth}
		\centering
		\includegraphics[width=\linewidth]{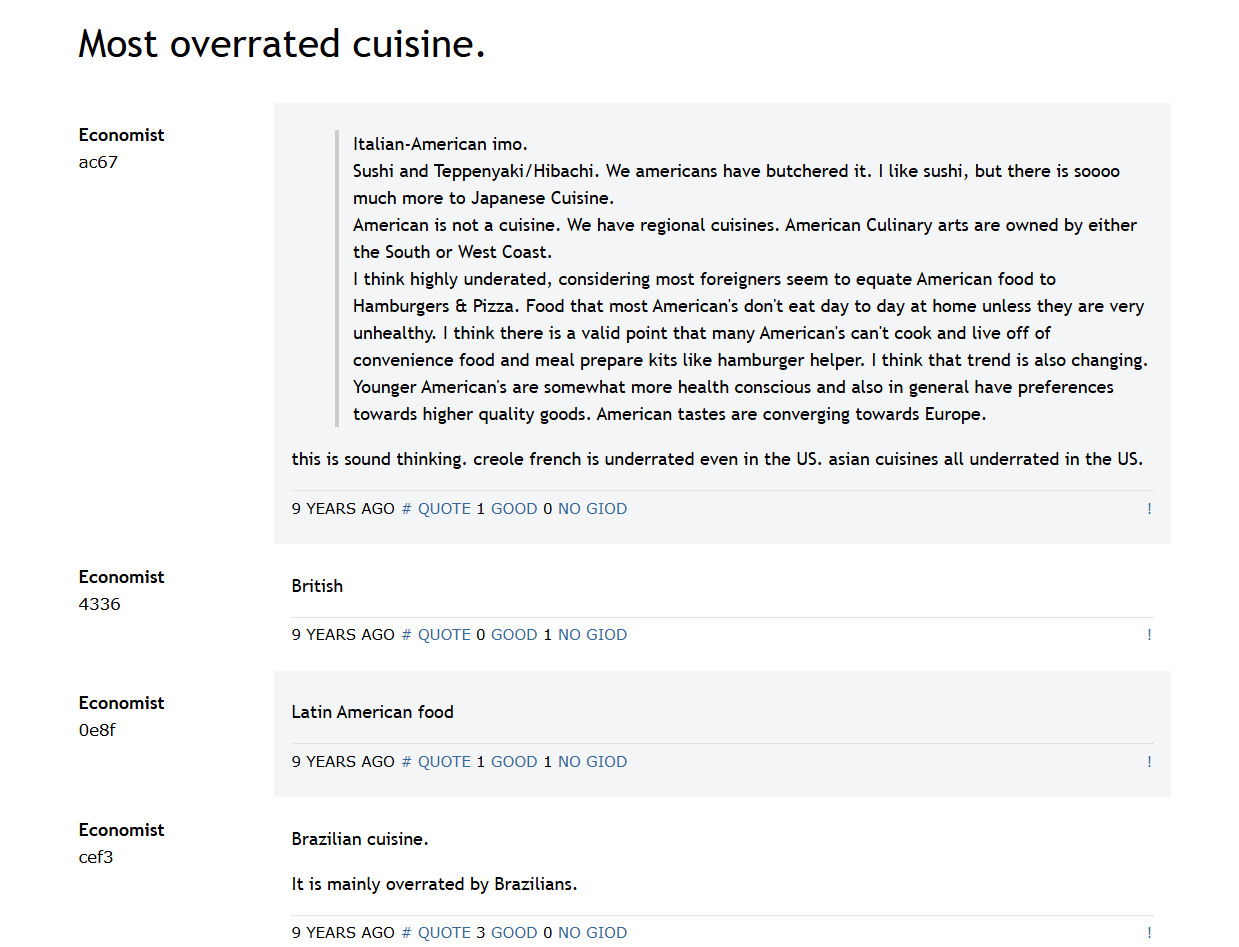}
		\caption*{\footnotesize Food}
	\end{minipage}\hfill
	\begin{minipage}[t]{0.31\textwidth}
		\centering
		\includegraphics[width=\linewidth]{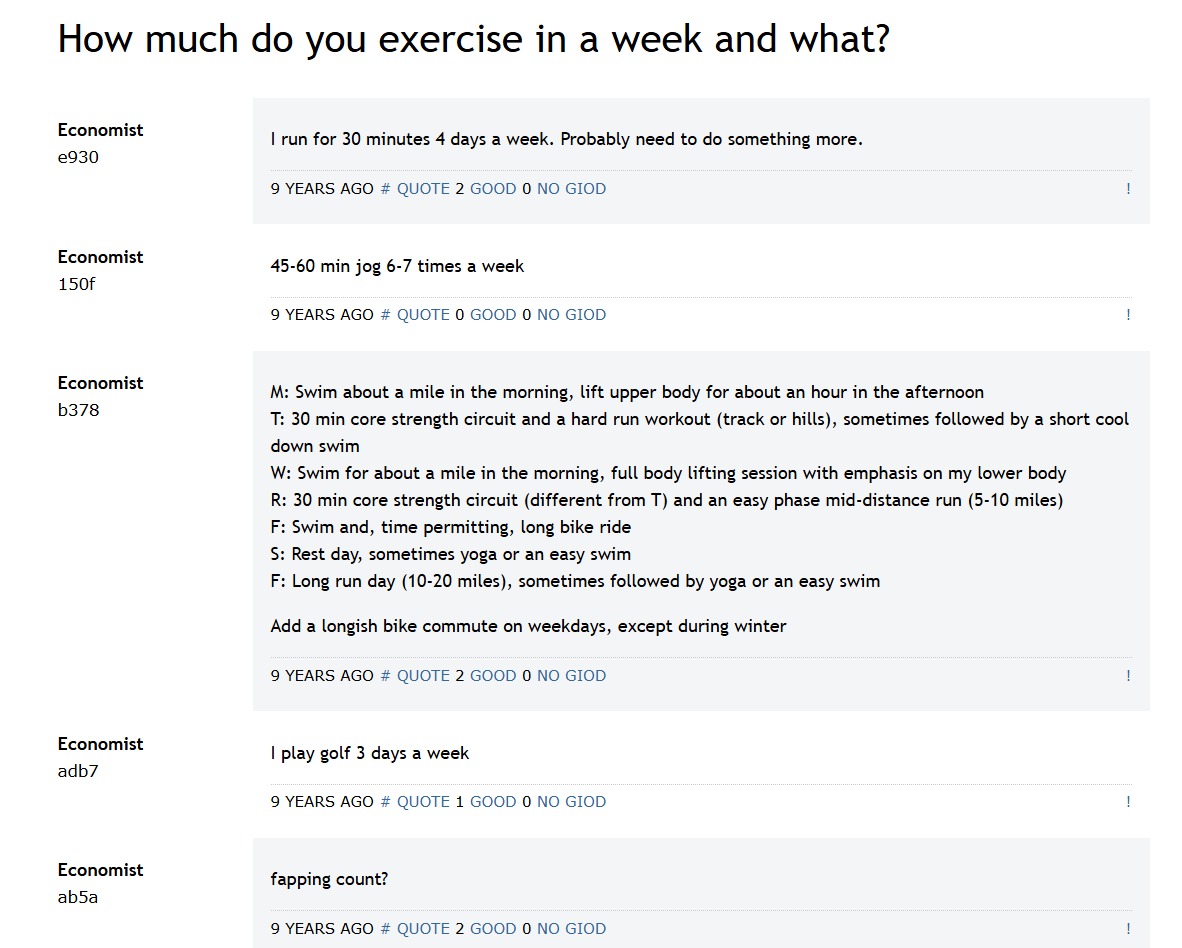}
		\caption*{\footnotesize Exercise}
	\end{minipage}\hfill
	\begin{minipage}[t]{0.31\textwidth}
		\centering
		\includegraphics[width=\linewidth]{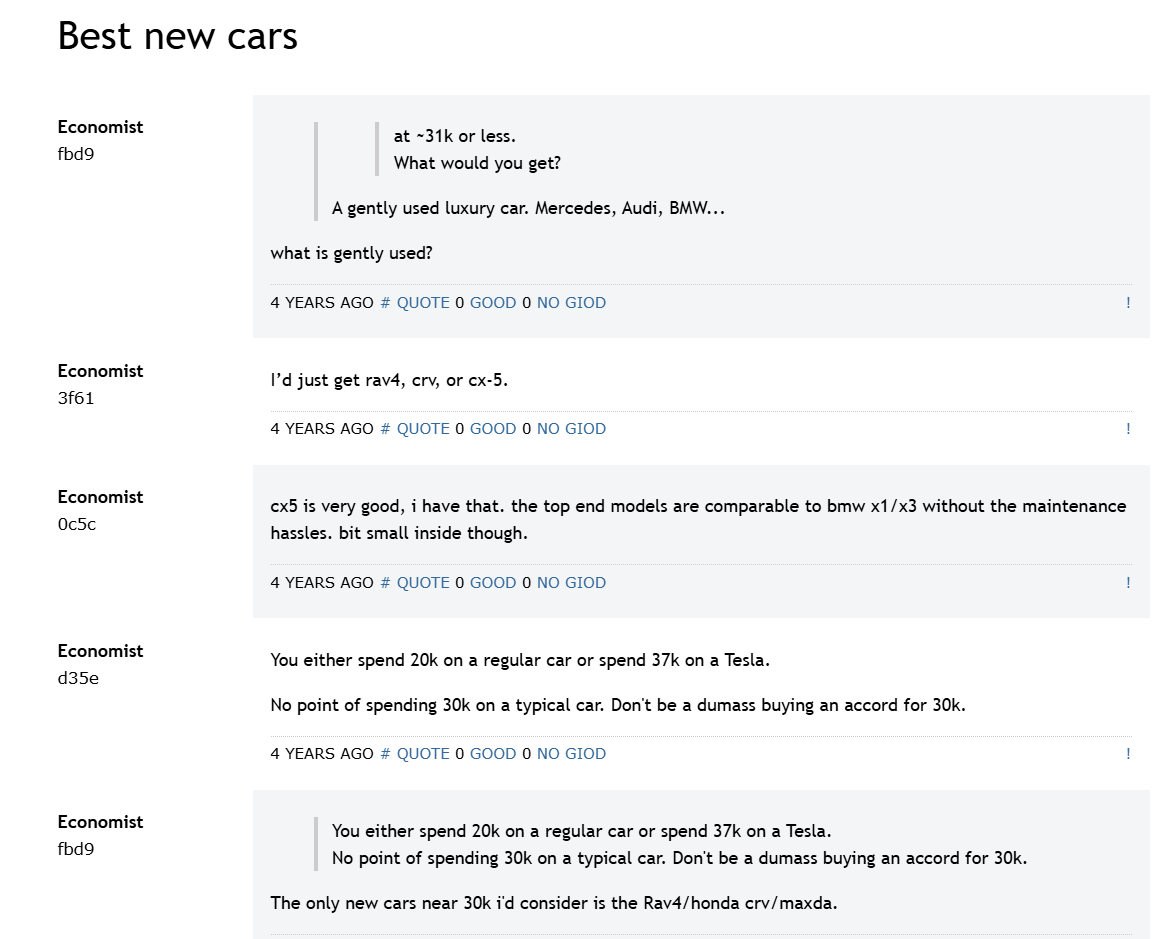}
		\caption*{\footnotesize Cars}
	\end{minipage}
	\caption{EJMR Response-Corpus Threads Outside Research Discussion}
	\label{fig:ejmr_nonresearch_examples}
\end{figure}

\subsection{Research Topics Trends}
\label{app:research_topic_trends}

\begin{figure}[H]
    \centering
    \includegraphics[width=0.9\textwidth]{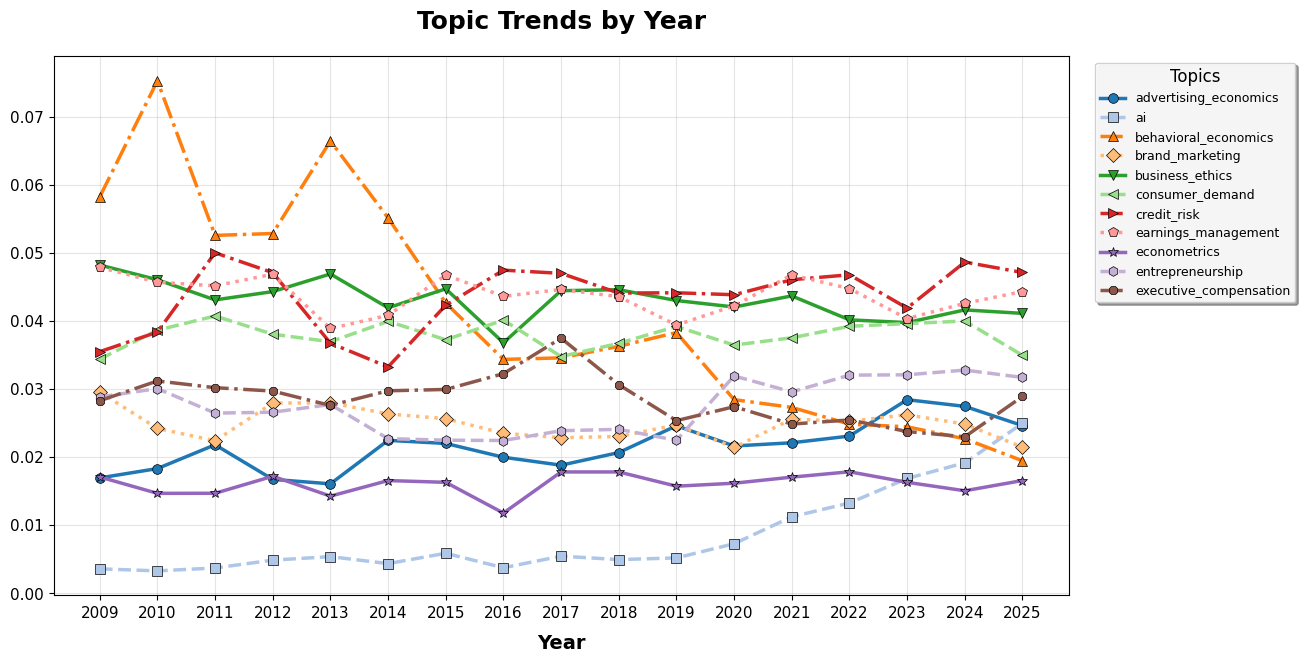}
    \caption{Research Topics Trend 1}
    \label{fig:research_topics_trend_1}
\end{figure}

\begin{figure}[H]
    \centering
    \includegraphics[width=0.9\textwidth]{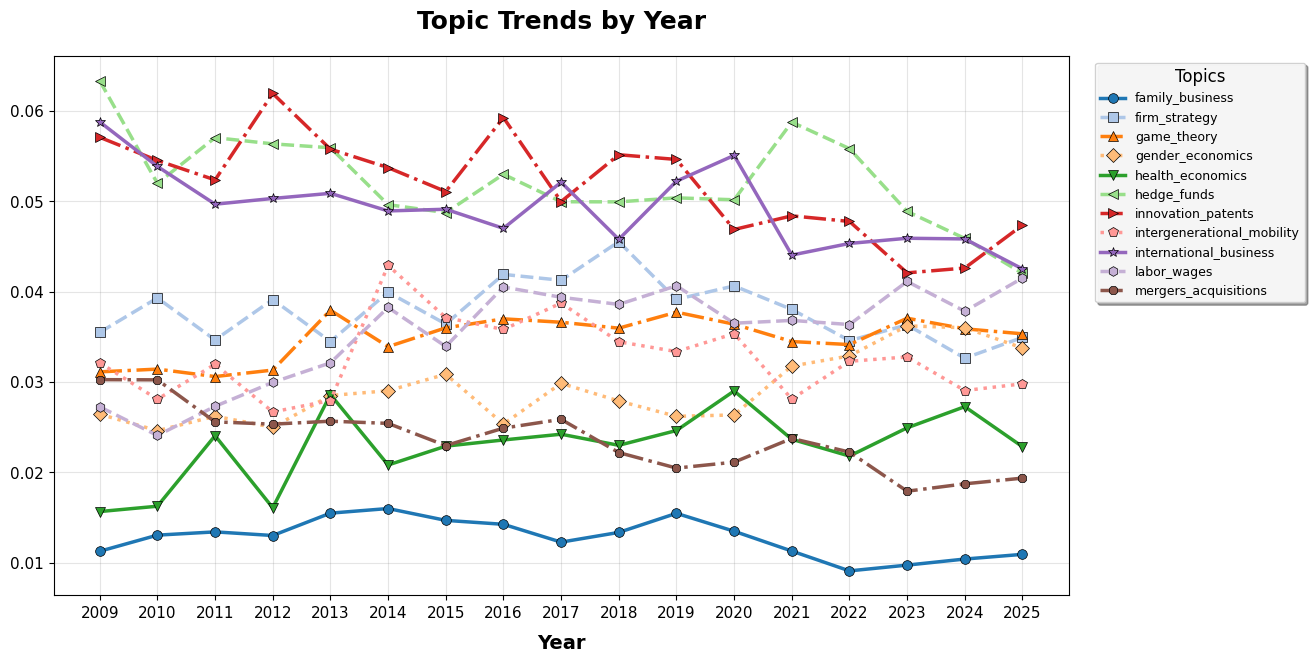}
    \caption{Research Topics Trend 2}
    \label{fig:research_topics_trend_2}
\end{figure}

\begin{figure}[H]
    \centering
    \includegraphics[width=0.9\textwidth]{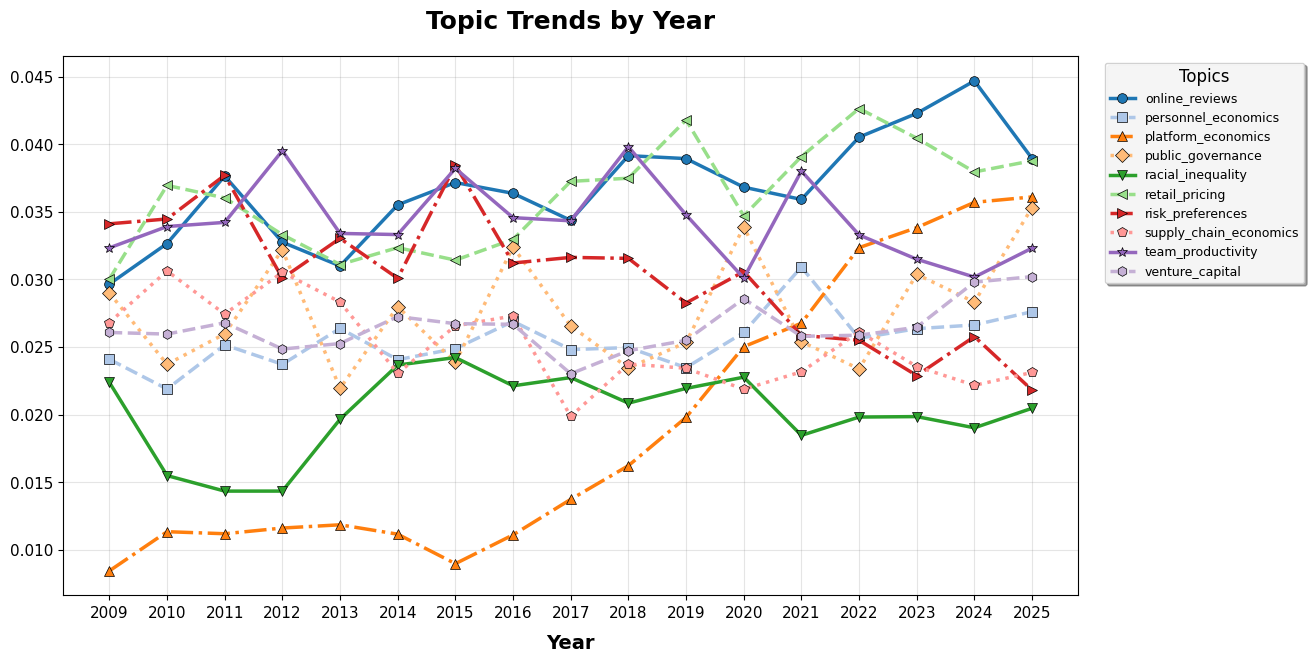}
    \caption{Research Topics Trend 3}
\label{fig:research_topics_trend_3}
\end{figure}

\subsection{Descriptive Properties of EJMR Response-Corpus Text Length}

\begin{figure}[H]
	\centering
	\includegraphics[width=0.48\textwidth]{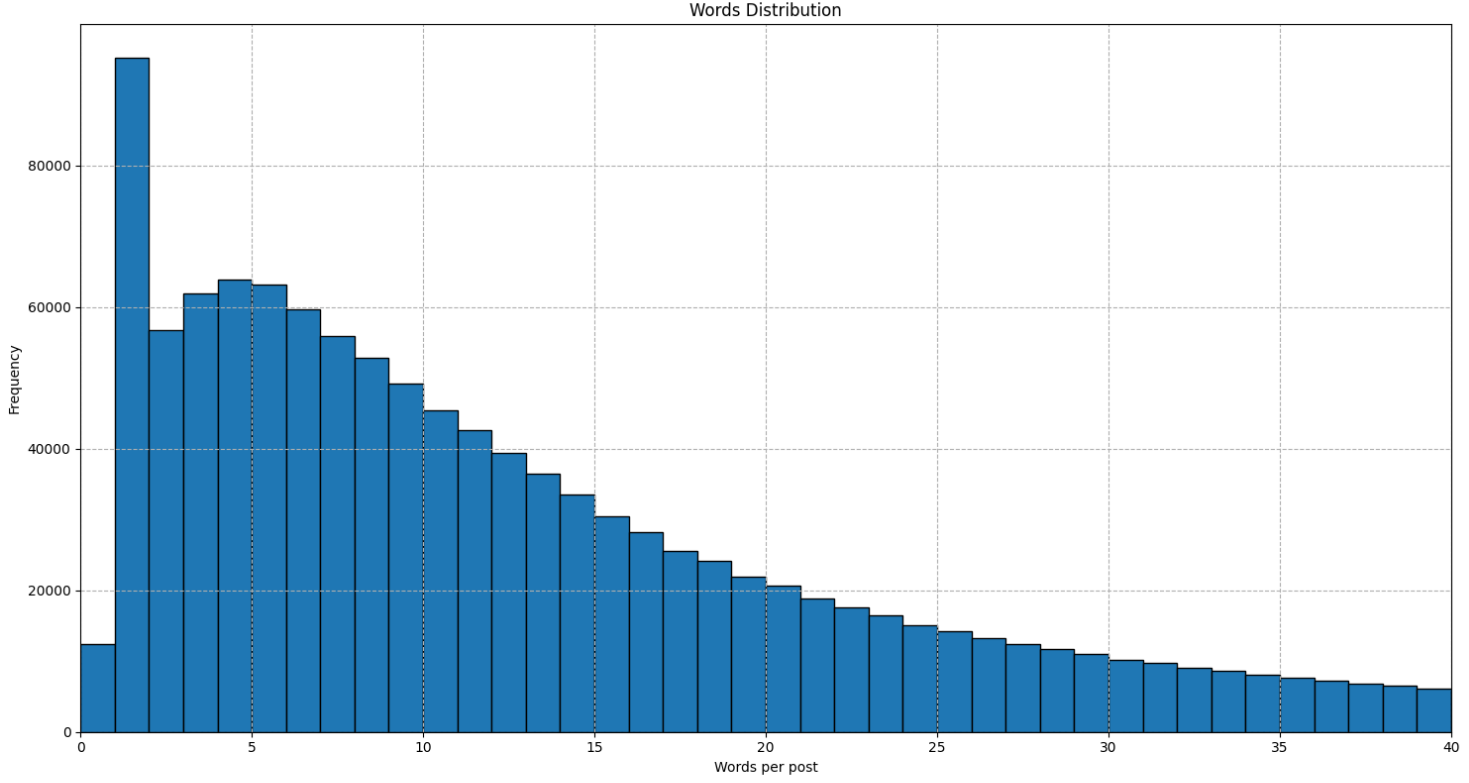}
	\hfill
	\includegraphics[width=0.48\textwidth]{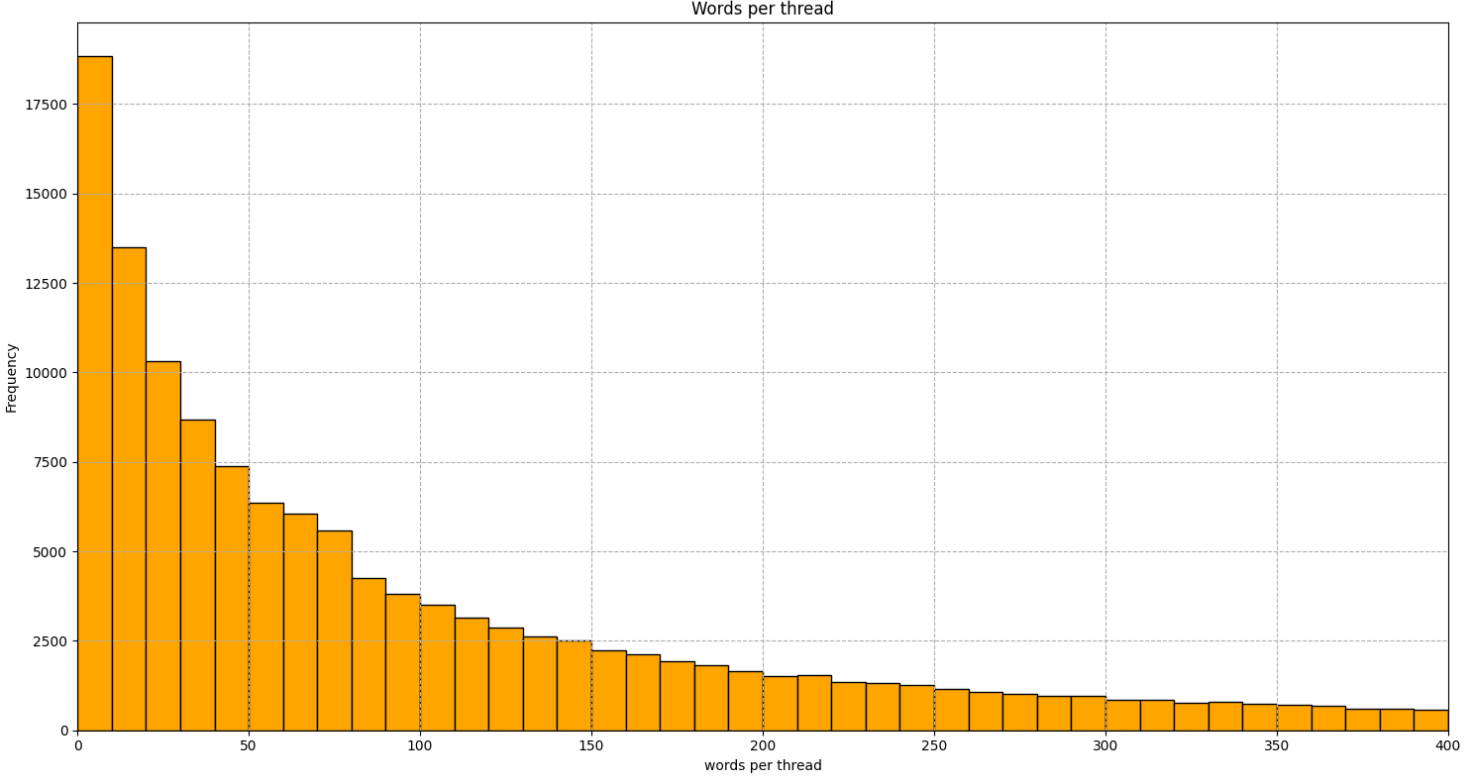}
	\caption{Descriptive Properties of EJMR Response-Corpus Text Length}
	\label{fig:text_length}
\end{figure}

Figure \ref{fig:text_length} provides additional context on the EJMR response corpus. Post-level text and thread-level text are both heterogeneous in length, which reinforces the value of semantic models over narrow keyword strategies. In a corpus where some responses are short, sarcastic, and highly compressed while others are longer and more explanatory, context-sensitive embeddings and response-level scoring are especially useful.

\section{Supplementary Regression and Descriptive Tables}
\subsection{Cross-Sectional Tables}
\label{app:static_effect_tables}
The tables below report cross-sectional topicality coefficients. Columns (1)--(4) progressively add forum indicators, year fixed effects, and thread-title emotion controls; standard errors are clustered by year and thread.

\begin{longtable}{lrrrr}
\caption{Cross-Sectional Results for Openness}\label{tab:openness_main_effects}\\
\toprule
 & (1) & (2) & (3) & (4) \\
\midrule
\endfirsthead
\toprule
 & (1) & (2) & (3) & (4) \\
\midrule
\endhead
Artificial Intelligence (AI) & -0.1684*** & -0.1713*** & -0.1209** & -0.1175** \\
 & (0.0471) & (0.0480) & (0.0522) & (0.0533) \\
Advertising Economics & -0.1402** & -0.1407** & -0.1151* & -0.0684 \\
 & (0.0687) & (0.0647) & (0.0623) & (0.0601) \\
Behavioral Economics & -0.1577*** & -0.0796*** & -0.0889*** & -0.0557** \\
 & (0.0285) & (0.0231) & (0.0220) & (0.0222) \\
Brand Marketing & 0.3529** & 0.3525** & 0.3051** & 0.1997 \\
 & (0.1765) & (0.1629) & (0.1516) & (0.1497) \\
Business Ethics & -0.9133*** & -0.8067*** & -0.7477*** & -0.6927*** \\
 & (0.0854) & (0.0801) & (0.0714) & (0.0740) \\
Consumer Demand & 0.0376 & -0.0353 & -0.0720 & -0.0633 \\
 & (0.1333) & (0.1132) & (0.1169) & (0.1178) \\
Credit Risk & -0.0784 & -0.0606 & -0.0443 & -0.0292 \\
 & (0.0492) & (0.0425) & (0.0397) & (0.0387) \\
Earnings Management & -0.4031*** & -0.3224*** & -0.3947*** & -0.4494*** \\
 & (0.1084) & (0.1083) & (0.1004) & (0.1059) \\
Econometrics & 0.3260*** & -0.0217 & 0.0036 & 0.0414 \\
 & (0.0357) & (0.0452) & (0.0533) & (0.0531) \\
Entrepreneurship & 0.1966** & 0.1881** & 0.1935** & 0.1723** \\
 & (0.0862) & (0.0857) & (0.0813) & (0.0795) \\
Executive Compensation & 0.1210* & 0.0909 & 0.1633** & 0.1688** \\
 & (0.0666) & (0.0724) & (0.0739) & (0.0769) \\
Family Business & -0.2328*** & -0.2083*** & -0.2196*** & -0.1863*** \\
 & (0.0773) & (0.0698) & (0.0719) & (0.0717) \\
Firm Strategy & 0.3213*** & 0.1418* & 0.0903 & 0.1485* \\
 & (0.0748) & (0.0861) & (0.0799) & (0.0822) \\
Game Theory & 0.1234* & 0.1250* & 0.0855 & 0.0141 \\
 & (0.0647) & (0.0701) & (0.0530) & (0.0599) \\
Gender Economics & -0.7919*** & -0.7194*** & -0.7257*** & -0.7090*** \\
 & (0.0848) & (0.0893) & (0.0803) & (0.0828) \\
Health Economics & -0.3719*** & -0.3523*** & -0.3450*** & -0.3513*** \\
 & (0.0603) & (0.0540) & (0.0561) & (0.0552) \\
Hedge Funds & -0.0155 & 0.0703 & 0.1042 & 0.1022 \\
 & (0.0976) & (0.0861) & (0.0762) & (0.0776) \\
Innovation and Patents & -0.0013 & 0.1007* & 0.1074* & 0.0803 \\
 & (0.0565) & (0.0608) & (0.0607) & (0.0593) \\
Intergenerational Mobility & -0.1120 & 0.0055 & -0.0352 & -0.0727 \\
 & (0.0768) & (0.0755) & (0.0652) & (0.0630) \\
International Business & 0.0959* & 0.0695 & 0.0540 & 0.0412 \\
 & (0.0581) & (0.0521) & (0.0523) & (0.0500) \\
Labor and Wages & 0.4500*** & 0.4497*** & 0.4475*** & 0.4373*** \\
 & (0.0900) & (0.0955) & (0.0930) & (0.0957) \\
Mergers and Acquisitions & 0.3707*** & 0.2957*** & 0.3172*** & 0.3394*** \\
 & (0.0867) & (0.0792) & (0.0799) & (0.0803) \\
Online Reviews & 0.5497*** & 0.5088*** & 0.4867*** & 0.4344*** \\
 & (0.0744) & (0.0666) & (0.0611) & (0.0605) \\
Personnel Economics & 0.3443*** & 0.2645*** & 0.1873*** & 0.1626** \\
 & (0.0739) & (0.0794) & (0.0707) & (0.0709) \\
Platform Economics & -0.2399*** & -0.2457*** & -0.2226** & -0.1744** \\
 & (0.0803) & (0.0824) & (0.0882) & (0.0867) \\
Public Governance & 0.0577 & 0.0805 & 0.1112*** & 0.0958** \\
 & (0.0635) & (0.0592) & (0.0421) & (0.0408) \\
Racial Inequality & -0.3023*** & -0.2376*** & -0.1758*** & -0.1544*** \\
 & (0.0696) & (0.0775) & (0.0469) & (0.0470) \\
Retail Pricing & 0.2327** & 0.1867* & 0.2190** & 0.2858*** \\
 & (0.0986) & (0.1022) & (0.1000) & (0.1104) \\
Risk Preferences & 0.1878** & 0.2181*** & 0.2108*** & 0.2476*** \\
 & (0.0898) & (0.0825) & (0.0719) & (0.0769) \\
Supply Chain Economics & -0.2954*** & -0.2579** & -0.2486** & -0.2574** \\
 & (0.1135) & (0.1141) & (0.1080) & (0.1097) \\
Team Productivity & 0.2351*** & 0.2161*** & 0.2468*** & 0.2266*** \\
 & (0.0823) & (0.0580) & (0.0567) & (0.0580) \\
Venture Capital & -0.1648 & -0.1317 & -0.1533 & -0.1531 \\
 & (0.1263) & (0.1233) & (0.1118) & (0.1076) \\
\midrule
Forum Indicators & No & Yes & Yes & Yes \\
Year Fixed Effects & No & No & Yes & Yes \\
Thread-Title Emotion Controls & No & No & No & Yes \\
Observations & 1,255,942 & 1,255,942 & 1,255,942 & 1,255,942 \\
Pseudo $R^2$ & 0.0012 & 0.0017 & 0.0021 & 0.0022 \\
\bottomrule
\end{longtable}

\noindent\footnotesize Notes: Entries are regression coefficients, with standard errors in parentheses. The dependent variable is reply-level openness, scored on the $[0,1]$ scale. Columns (1)--(4) sequentially add forum indicators, year fixed effects, and thread-title emotion controls. Standard errors are two-way clustered by calendar year and thread. * $p<0.1$, ** $p<0.05$, *** $p<0.01$.
\normalsize

\begin{longtable}{lrrrr}
\caption{Cross-Sectional Results for Curiosity}\label{tab:curiosity_main_effects}\\
\toprule
 & (1) & (2) & (3) & (4) \\
\midrule
\endfirsthead
\toprule
 & (1) & (2) & (3) & (4) \\
\midrule
\endhead
Artificial Intelligence (AI) & -0.1834*** & -0.1859*** & -0.1519*** & -0.1529*** \\
 & (0.0433) & (0.0425) & (0.0489) & (0.0498) \\
Advertising Economics & 0.1153** & 0.1257** & 0.1525*** & 0.1322*** \\
 & (0.0569) & (0.0557) & (0.0499) & (0.0466) \\
Behavioral Economics & 0.1146*** & 0.1719*** & 0.1702*** & 0.0784*** \\
 & (0.0237) & (0.0217) & (0.0189) & (0.0196) \\
Brand Marketing & 0.1988 & 0.1776 & 0.1447 & 0.1528 \\
 & (0.1351) & (0.1233) & (0.1120) & (0.1128) \\
Business Ethics & -0.6759*** & -0.5910*** & -0.5482*** & -0.5901*** \\
 & (0.0998) & (0.0939) & (0.0863) & (0.0883) \\
Consumer Demand & -0.2726** & -0.2691** & -0.2912*** & -0.2800*** \\
 & (0.1090) & (0.1097) & (0.1105) & (0.1068) \\
Credit Risk & 0.0759** & 0.0850** & 0.0970*** & 0.0624* \\
 & (0.0383) & (0.0331) & (0.0335) & (0.0331) \\
Earnings Management & -0.3319*** & -0.2368*** & -0.2922*** & -0.2756*** \\
 & (0.0865) & (0.0885) & (0.0903) & (0.0957) \\
Econometrics & 0.3394*** & 0.0531 & 0.0810 & 0.0421 \\
 & (0.0377) & (0.0436) & (0.0523) & (0.0537) \\
Entrepreneurship & 0.1621** & 0.1625** & 0.1671** & 0.1757*** \\
 & (0.0653) & (0.0687) & (0.0682) & (0.0662) \\
Executive Compensation & 0.1631** & 0.1369* & 0.2003** & 0.1806** \\
 & (0.0778) & (0.0784) & (0.0835) & (0.0805) \\
Family Business & -0.1814** & -0.1830*** & -0.1980*** & -0.1695** \\
 & (0.0747) & (0.0650) & (0.0673) & (0.0700) \\
Firm Strategy & 0.1766*** & -0.0015 & -0.0487 & -0.0263 \\
 & (0.0675) & (0.0808) & (0.0743) & (0.0718) \\
Game Theory & 0.2795*** & 0.2781*** & 0.2590*** & 0.1625*** \\
 & (0.0543) & (0.0622) & (0.0583) & (0.0602) \\
Gender Economics & -0.7902*** & -0.7120*** & -0.7161*** & -0.7405*** \\
 & (0.0685) & (0.0724) & (0.0683) & (0.0674) \\
Health Economics & -0.2151*** & -0.2207*** & -0.2197*** & -0.2161*** \\
 & (0.0500) & (0.0420) & (0.0412) & (0.0430) \\
Hedge Funds & -0.0624 & -0.0111 & 0.0076 & -0.0199 \\
 & (0.0783) & (0.0746) & (0.0645) & (0.0691) \\
Innovation and Patents & 0.1745*** & 0.2600*** & 0.2654*** & 0.2212*** \\
 & (0.0588) & (0.0607) & (0.0603) & (0.0585) \\
Intergenerational Mobility & -0.2514*** & -0.1441** & -0.1766*** & -0.2159*** \\
 & (0.0658) & (0.0632) & (0.0564) & (0.0544) \\
International Business & 0.2548*** & 0.2152*** & 0.2029*** & 0.1426** \\
 & (0.0735) & (0.0665) & (0.0666) & (0.0647) \\
Labor and Wages & 0.2188** & 0.2093** & 0.2038** & 0.2587*** \\
 & (0.0850) & (0.0887) & (0.0906) & (0.0880) \\
Mergers and Acquisitions & 0.3169*** & 0.2555*** & 0.2666*** & 0.2887*** \\
 & (0.0812) & (0.0805) & (0.0802) & (0.0813) \\
Online Reviews & 0.3963*** & 0.3455*** & 0.3307*** & 0.3183*** \\
 & (0.0775) & (0.0678) & (0.0632) & (0.0623) \\
Personnel Economics & 0.3774*** & 0.3175*** & 0.2617** & 0.2222** \\
 & (0.0916) & (0.1067) & (0.1059) & (0.0993) \\
Platform Economics & -0.1845** & -0.1809** & -0.1685** & -0.1286 \\
 & (0.0856) & (0.0836) & (0.0854) & (0.0794) \\
Public Governance & -0.0151 & 0.0143 & 0.0387 & 0.0681 \\
 & (0.0707) & (0.0685) & (0.0546) & (0.0536) \\
Racial Inequality & 0.0253 & 0.0746 & 0.1242** & 0.1667*** \\
 & (0.0673) & (0.0744) & (0.0565) & (0.0575) \\
Retail Pricing & 0.3390*** & 0.2340** & 0.2448** & 0.2333** \\
 & (0.1038) & (0.1125) & (0.1138) & (0.1156) \\
Risk Preferences & 0.0367 & 0.0563 & 0.0439 & 0.1725* \\
 & (0.0978) & (0.0920) & (0.0837) & (0.0882) \\
Supply Chain Economics & -0.3692*** & -0.3040*** & -0.2852*** & -0.2592** \\
 & (0.1094) & (0.1132) & (0.1074) & (0.1059) \\
Team Productivity & 0.0653 & 0.0574 & 0.0827 & 0.1515*** \\
 & (0.0750) & (0.0620) & (0.0603) & (0.0584) \\
Venture Capital & -0.1788** & -0.1557** & -0.1724** & -0.1752*** \\
 & (0.0750) & (0.0741) & (0.0670) & (0.0675) \\
\midrule
Forum Indicators & No & Yes & Yes & Yes \\
Year Fixed Effects & No & No & Yes & Yes \\
Thread-Title Emotion Controls & No & No & No & Yes \\
Observations & 1,255,942 & 1,255,942 & 1,255,942 & 1,255,942 \\
Pseudo $R^2$ & 0.0010 & 0.0014 & 0.0017 & 0.0018 \\
\bottomrule
\end{longtable}

\noindent\footnotesize Notes: Entries are regression coefficients, with standard errors in parentheses. The dependent variable is reply-level curiosity, scored on the $[0,1]$ scale. Columns (1)--(4) sequentially add forum indicators, year fixed effects, and thread-title emotion controls. Standard errors are two-way clustered by calendar year and thread. * $p<0.1$, ** $p<0.05$, *** $p<0.01$.
\normalsize

\begin{longtable}{lrrrr}
\caption{Cross-Sectional Results for Negative Tone}\label{tab:negative_main_effects}\\
\toprule
 & (1) & (2) & (3) & (4) \\
\midrule
\endfirsthead
\toprule
 & (1) & (2) & (3) & (4) \\
\midrule
\endhead
Artificial Intelligence (AI) & 0.1629*** & 0.1732*** & 0.1092* & 0.1248** \\
 & (0.0515) & (0.0546) & (0.0560) & (0.0536) \\
Advertising Economics & -0.2153** & -0.2061*** & -0.2344*** & -0.2099*** \\
 & (0.0847) & (0.0754) & (0.0707) & (0.0649) \\
Behavioral Economics & 0.1794*** & 0.0741*** & 0.0848*** & -0.0118 \\
 & (0.0293) & (0.0239) & (0.0229) & (0.0204) \\
Brand Marketing & -0.5974*** & -0.6015*** & -0.5154*** & -0.4141*** \\
 & (0.1929) & (0.1839) & (0.1681) & (0.1582) \\
Business Ethics & 1.7327*** & 1.5881*** & 1.4950*** & 1.2873*** \\
 & (0.1274) & (0.1208) & (0.1066) & (0.1044) \\
Consumer Demand & -0.3906*** & -0.3099*** & -0.3106*** & -0.0440 \\
 & (0.1186) & (0.1057) & (0.1146) & (0.1156) \\
Credit Risk & 0.2681*** & 0.2422*** & 0.2156*** & 0.1458*** \\
 & (0.0633) & (0.0539) & (0.0473) & (0.0449) \\
Earnings Management & 0.4535*** & 0.3444*** & 0.4002*** & 0.4912*** \\
 & (0.1101) & (0.1129) & (0.0981) & (0.1084) \\
Econometrics & -0.5238*** & -0.0740 & -0.1027 & -0.1228* \\
 & (0.0501) & (0.0574) & (0.0659) & (0.0632) \\
Entrepreneurship & -0.0949 & -0.0903 & -0.1119 & -0.0644 \\
 & (0.1028) & (0.1025) & (0.1025) & (0.1019) \\
Executive Compensation & -0.1821** & -0.1590 & -0.2120** & -0.2104** \\
 & (0.0888) & (0.1019) & (0.0919) & (0.0970) \\
Family Business & 0.3325*** & 0.3008*** & 0.3046*** & 0.2709*** \\
 & (0.0929) & (0.0854) & (0.0867) & (0.0852) \\
Firm Strategy & -0.5687*** & -0.3669*** & -0.2894*** & -0.3336*** \\
 & (0.0806) & (0.0995) & (0.0854) & (0.0865) \\
Game Theory & -0.3733*** & -0.3717*** & -0.3297*** & -0.1086 \\
 & (0.0784) & (0.0837) & (0.0600) & (0.0724) \\
Gender Economics & 0.9224*** & 0.8354*** & 0.8258*** & 0.7773*** \\
 & (0.1070) & (0.1113) & (0.1008) & (0.1019) \\
Health Economics & 0.6491*** & 0.6096*** & 0.6003*** & 0.5325*** \\
 & (0.0592) & (0.0541) & (0.0562) & (0.0577) \\
Hedge Funds & 0.1135 & 0.0236 & -0.0161 & 0.0027 \\
 & (0.1016) & (0.0889) & (0.0678) & (0.0703) \\
Innovation and Patents & 0.3977*** & 0.2713*** & 0.2620*** & 0.1902*** \\
 & (0.0711) & (0.0710) & (0.0704) & (0.0703) \\
Intergenerational Mobility & -0.1783** & -0.3067*** & -0.2579*** & -0.1383** \\
 & (0.0857) & (0.0924) & (0.0745) & (0.0700) \\
International Business & -0.3167*** & -0.2909*** & -0.2750*** & -0.2166*** \\
 & (0.0657) & (0.0615) & (0.0615) & (0.0599) \\
Labor and Wages & -0.1416 & -0.1669* & -0.1448 & -0.3064*** \\
 & (0.0987) & (0.0961) & (0.0899) & (0.0920) \\
Mergers and Acquisitions & -0.1521 & -0.0550 & -0.0841 & -0.1444 \\
 & (0.1059) & (0.0949) & (0.0944) & (0.0906) \\
Online Reviews & -0.2780*** & -0.2352*** & -0.2040*** & -0.2309*** \\
 & (0.0677) & (0.0541) & (0.0456) & (0.0446) \\
Personnel Economics & -0.9562*** & -0.8280*** & -0.7419*** & -0.4832*** \\
 & (0.1158) & (0.1032) & (0.0538) & (0.0586) \\
Platform Economics & 0.6434*** & 0.6501*** & 0.6165*** & 0.4856*** \\
 & (0.0797) & (0.0781) & (0.0886) & (0.0891) \\
Public Governance & -0.0959 & -0.1233 & -0.1632** & -0.1348** \\
 & (0.0969) & (0.0922) & (0.0701) & (0.0641) \\
Racial Inequality & 0.5786*** & 0.5005*** & 0.4322*** & 0.4202*** \\
 & (0.0825) & (0.0943) & (0.0552) & (0.0590) \\
Retail Pricing & -0.0660 & -0.0154 & -0.0191 & -0.2660** \\
 & (0.1042) & (0.1053) & (0.1050) & (0.1260) \\
Risk Preferences & -0.0207 & -0.0542 & -0.0259 & -0.2647*** \\
 & (0.0856) & (0.0696) & (0.0608) & (0.0711) \\
Supply Chain Economics & 0.0578 & 0.0259 & 0.0057 & 0.0841 \\
 & (0.1234) & (0.1240) & (0.1200) & (0.1202) \\
Team Productivity & -0.3506*** & -0.3137*** & -0.3435*** & -0.3101*** \\
 & (0.1133) & (0.0835) & (0.0738) & (0.0712) \\
Venture Capital & -0.1741 & -0.2048 & -0.1733 & -0.0927 \\
 & (0.1418) & (0.1349) & (0.1251) & (0.1207) \\
\midrule
Forum Indicators & No & Yes & Yes & Yes \\
Year Fixed Effects & No & No & Yes & Yes \\
Thread-Title Emotion Controls & No & No & No & Yes \\
Observations & 1,255,942 & 1,255,942 & 1,255,942 & 1,255,942 \\
Pseudo $R^2$ & 0.0026 & 0.0034 & 0.0039 & 0.0043 \\
\bottomrule
\end{longtable}

\noindent\footnotesize Notes: Entries are regression coefficients, with standard errors in parentheses. The dependent variable is reply-level negative tone, scored on the $[0,1]$ scale. Columns (1)--(4) sequentially add forum indicators, year fixed effects, and thread-title emotion controls. Standard errors are two-way clustered by calendar year and thread. * $p<0.1$, ** $p<0.05$, *** $p<0.01$.
\normalsize

\begin{longtable}{lrrrr}
\caption{Cross-Sectional Results for Poisonousness}\label{tab:poisonous_main_effects}\\
\toprule
 & (1) & (2) & (3) & (4) \\
\midrule
\endfirsthead
\toprule
 & (1) & (2) & (3) & (4) \\
\midrule
\endhead
Artificial Intelligence (AI) & -0.0690 & -0.0672 & -0.1169* & -0.1067* \\
 & (0.0569) & (0.0606) & (0.0644) & (0.0635) \\
Advertising Economics & 0.0618 & 0.0861 & 0.0569 & 0.0581 \\
 & (0.0866) & (0.0786) & (0.0746) & (0.0720) \\
Behavioral Economics & 0.3302*** & 0.2129*** & 0.2151*** & 0.1920*** \\
 & (0.0334) & (0.0302) & (0.0315) & (0.0298) \\
Brand Marketing & -0.4916** & -0.5366** & -0.4643** & -0.3972* \\
 & (0.2437) & (0.2265) & (0.2141) & (0.2073) \\
Business Ethics & 1.4134*** & 1.2593*** & 1.1817*** & 1.0801*** \\
 & (0.1454) & (0.1360) & (0.1217) & (0.1239) \\
Consumer Demand & -0.2807** & -0.1260 & -0.1385 & -0.0006 \\
 & (0.1303) & (0.1052) & (0.1171) & (0.1213) \\
Credit Risk & 0.1605** & 0.1324** & 0.1089** & 0.0834 \\
 & (0.0675) & (0.0576) & (0.0530) & (0.0534) \\
Earnings Management & 0.6159*** & 0.5298*** & 0.5926*** & 0.6473*** \\
 & (0.1179) & (0.1230) & (0.1040) & (0.1138) \\
Econometrics & -0.6365*** & -0.1271** & -0.1629*** & -0.1642*** \\
 & (0.0384) & (0.0533) & (0.0626) & (0.0628) \\
Entrepreneurship & 0.1011 & 0.1105 & 0.0843 & 0.1040 \\
 & (0.0988) & (0.0968) & (0.0955) & (0.0934) \\
Executive Compensation & 0.0688 & 0.0935 & 0.0260 & 0.0271 \\
 & (0.0854) & (0.1035) & (0.0953) & (0.0967) \\
Family Business & 0.2096*** & 0.1633** & 0.1769** & 0.1460* \\
 & (0.0811) & (0.0758) & (0.0808) & (0.0820) \\
Firm Strategy & -1.1077*** & -0.8501*** & -0.7789*** & -0.8237*** \\
 & (0.0732) & (0.0854) & (0.0729) & (0.0756) \\
Game Theory & -0.0601 & -0.0471 & -0.0208 & 0.1323 \\
 & (0.1037) & (0.1023) & (0.0809) & (0.0855) \\
Gender Economics & 1.0423*** & 0.9611*** & 0.9555*** & 0.9401*** \\
 & (0.1191) & (0.1175) & (0.1096) & (0.1106) \\
Health Economics & 0.5587*** & 0.5136*** & 0.5097*** & 0.4789*** \\
 & (0.0602) & (0.0551) & (0.0568) & (0.0575) \\
Hedge Funds & 0.2064 & 0.0909 & 0.0422 & 0.0641 \\
 & (0.1265) & (0.1145) & (0.0995) & (0.1029) \\
Innovation and Patents & 0.5171*** & 0.3677*** & 0.3548*** & 0.3406*** \\
 & (0.0505) & (0.0570) & (0.0571) & (0.0579) \\
Intergenerational Mobility & -0.0678 & -0.2297** & -0.1818** & -0.1149 \\
 & (0.0886) & (0.0897) & (0.0770) & (0.0743) \\
International Business & -0.1897** & -0.1689** & -0.1440** & -0.1041 \\
 & (0.0819) & (0.0706) & (0.0688) & (0.0654) \\
Labor and Wages & -0.4772*** & -0.4989*** & -0.4820*** & -0.5700*** \\
 & (0.1119) & (0.1160) & (0.1106) & (0.1124) \\
Mergers and Acquisitions & -0.4041*** & -0.3052*** & -0.3279*** & -0.3737*** \\
 & (0.1162) & (0.1038) & (0.1038) & (0.1013) \\
Online Reviews & -0.4231*** & -0.3857*** & -0.3609*** & -0.3616*** \\
 & (0.0791) & (0.0674) & (0.0643) & (0.0640) \\
Personnel Economics & -0.5500*** & -0.4219*** & -0.3448*** & -0.1993*** \\
 & (0.1163) & (0.1001) & (0.0654) & (0.0712) \\
Platform Economics & 0.4544*** & 0.4380*** & 0.4082*** & 0.3328*** \\
 & (0.0719) & (0.0727) & (0.0824) & (0.0826) \\
Public Governance & -0.1430* & -0.1783** & -0.2143*** & -0.2034*** \\
 & (0.0844) & (0.0726) & (0.0579) & (0.0528) \\
Racial Inequality & 0.6901*** & 0.6006*** & 0.5361*** & 0.5214*** \\
 & (0.0790) & (0.0832) & (0.0477) & (0.0496) \\
Retail Pricing & -0.4580*** & -0.4387*** & -0.4188*** & -0.5658*** \\
 & (0.1127) & (0.1077) & (0.1088) & (0.1254) \\
Risk Preferences & -0.1565** & -0.2259*** & -0.1901*** & -0.3558*** \\
 & (0.0779) & (0.0719) & (0.0655) & (0.0726) \\
Supply Chain Economics & 0.5389*** & 0.5219*** & 0.4847*** & 0.5233*** \\
 & (0.1208) & (0.1183) & (0.1163) & (0.1161) \\
Team Productivity & -0.4232*** & -0.3790*** & -0.4071*** & -0.4067*** \\
 & (0.0992) & (0.0713) & (0.0693) & (0.0669) \\
Venture Capital & -0.2854** & -0.3002** & -0.2526** & -0.2074* \\
 & (0.1315) & (0.1299) & (0.1193) & (0.1134) \\
\midrule
Forum Indicators & No & Yes & Yes & Yes \\
Year Fixed Effects & No & No & Yes & Yes \\
Thread-Title Emotion Controls & No & No & No & Yes \\
Observations & 1,255,942 & 1,255,942 & 1,255,942 & 1,255,942 \\
Pseudo $R^2$ & 0.0026 & 0.0036 & 0.0041 & 0.0043 \\
\bottomrule
\end{longtable}

\noindent\footnotesize Notes: Entries are regression coefficients, with standard errors in parentheses. The dependent variable is reply-level poisonousness, scored on the $[0,1]$ scale. Columns (1)--(4) sequentially add forum indicators, year fixed effects, and thread-title emotion controls. Standard errors are two-way clustered by calendar year and thread. * $p<0.1$, ** $p<0.05$, *** $p<0.01$.
\normalsize

\subsection{Detailed Regression Results}
\label{app:detailed_regression_results}

The complete dynamic-regression estimates corresponding to the figures in Section \ref{sec:model} are reported below.

\begin{longtable}{lrrrr}
\caption{Dynamic Regression Results}\label{tab:dynamic_ai_full}\\
\toprule
 & Openness & Curiosity & Negative Tone & Poisonousness  \\
\midrule
\endfirsthead
\toprule
 & Openness & Curiosity & Negative Tone & Poisonousness  \\
\midrule
\endhead
\multicolumn{5}{l}{\textit{AI Topicality effect}}\\
Artificial Intelligence (AI) & -0.0544 & -0.1263*** & 0.0791* & -0.1088***  \\
 & (0.0397) & (0.0448) & (0.0404) & (0.0370)  \\
\addlinespace
\multicolumn{5}{l}{\textit{AI Topicality $\times$ year}}\\
AI Topicality $\times$ Year 2013 & -0.0099 & -0.0022 & 0.0293*** & 0.0007  \\
 & (0.0112) & (0.0198) & (0.0100) & (0.0082)  \\
AI Topicality $\times$ Year 2014 & -0.1070*** & -0.0013 & 0.1233*** & 0.0831***  \\
 & (0.0103) & (0.0192) & (0.0094) & (0.0080)  \\
AI Topicality $\times$ Year 2015 & -0.1318*** & -0.0732*** & 0.1727*** & 0.1471***  \\
 & (0.0104) & (0.0184) & (0.0098) & (0.0062)  \\
AI Topicality $\times$ Year 2016 & -0.1820*** & -0.0716*** & 0.2180*** & 0.1689***  \\
 & (0.0103) & (0.0198) & (0.0100) & (0.0066)  \\
AI Topicality $\times$ Year 2017 & -0.2340*** & -0.1261*** & 0.2610*** & 0.2360***  \\
 & (0.0101) & (0.0198) & (0.0091) & (0.0053)  \\
AI Topicality $\times$ Year 2018 & -0.1771*** & -0.1038*** & 0.2013*** & 0.1513***  \\
 & (0.0098) & (0.0195) & (0.0078) & (0.0039)  \\
AI Topicality $\times$ Year 2019 & -0.2019*** & -0.1302*** & 0.2140*** & 0.1830***  \\
 & (0.0087) & (0.0182) & (0.0069) & (0.0041)  \\
AI Topicality $\times$ Year 2020 & -0.1521*** & -0.1392*** & 0.1120*** & 0.0626***  \\
 & (0.0097) & (0.0191) & (0.0097) & (0.0065)  \\
AI Topicality $\times$ Year 2021 & -0.0536*** & -0.0770*** & 0.0078 & -0.0606***  \\
 & (0.0123) & (0.0206) & (0.0131) & (0.0098)  \\
AI Topicality $\times$ Year 2022 & 0.0275** & -0.0027 & -0.0587*** & -0.1602***  \\
 & (0.0139) & (0.0217) & (0.0128) & (0.0086)  \\
AI Topicality $\times$ Year 2023 & 0.0308** & 0.0857*** & -0.0326** & -0.0505***  \\
 & (0.0149) & (0.0222) & (0.0147) & (0.0111)  \\
AI Topicality $\times$ Year 2024 & 0.0133 & 0.0429* & -0.1036*** & -0.1320***  \\
 & (0.0169) & (0.0239) & (0.0190) & (0.0142)  \\
\midrule
Topicality Controls & Yes & Yes & Yes & Yes  \\
Forum Indicators & Yes & Yes & Yes & Yes  \\
Year Fixed Effects & Yes & Yes & Yes & Yes  \\
Thread-Title Emotion Controls & Yes & Yes & Yes & Yes  \\
Observations & 1,255,942 & 1,255,942 & 1,255,942 & 1,255,942  \\
Pseudo $R^2$ & 0.0022 & 0.0018 & 0.0044 & 0.0043  \\
\bottomrule
\end{longtable}

\noindent\footnotesize Notes: Dependent variables are reply-level sentiment scores on the $[0,1]$ scale. The first two columns measure openness and curiosity, for which higher values indicate greater openness and curiosity; the second two columns measure negative tone and poisonousness, for which higher values indicate more negative and poisonous language. Standard errors are two-way clustered by calendar year and thread. * $p<0.1$, ** $p<0.05$, *** $p<0.01$.
\normalsize

\begin{longtable}{lrrrr}
\caption{Other Topic Effect in Dynamic Regression}\label{tab:dynamic_other_topics}\\
\toprule
 & Openness & Curiosity & Negative Tone & Poisonousness \\
\midrule
\endfirsthead
\toprule
 & Openness & Curiosity & Negative Tone & Poisonousness \\
\midrule
\endhead
\multicolumn{5}{l}{\textit{AI Topicality effect}}\\
Artificial Intelligence (AI) & -0.0544 & -0.1263*** & 0.0791* & -0.1088*** \\
 & (0.0397) & (0.0448) & (0.0404) & (0.0370) \\
\addlinespace
\multicolumn{5}{l}{\textit{Other topicality effects}}\\
Advertising Economics & -0.0589 & 0.1388*** & -0.2209*** & 0.0444 \\
 & (0.0608) & (0.0471) & (0.0653) & (0.0720) \\
Behavioral Economics & -0.0539** & 0.0788*** & -0.0137 & 0.1873*** \\
 & (0.0227) & (0.0201) & (0.0208) & (0.0302) \\
Brand Marketing & 0.1860 & 0.1460 & -0.4056*** & -0.3676* \\
 & (0.1501) & (0.1136) & (0.1573) & (0.2073) \\
Business Ethics & -0.6966*** & -0.5944*** & 1.2982*** & 1.0806*** \\
 & (0.0747) & (0.0878) & (0.1050) & (0.1252) \\
Consumer Demand & -0.0401 & -0.2663** & -0.0438 & -0.0526 \\
 & (0.1178) & (0.1073) & (0.1102) & (0.1210) \\
Credit Risk & -0.0311 & 0.0601* & 0.1478*** & 0.0848 \\
 & (0.0387) & (0.0330) & (0.0446) & (0.0535) \\
Earnings Management & -0.4370*** & -0.2653*** & 0.4883*** & 0.6310*** \\
 & (0.1080) & (0.0969) & (0.1112) & (0.1162) \\
Econometrics & 0.0429 & 0.0424 & -0.1223* & -0.1677*** \\
 & (0.0534) & (0.0538) & (0.0630) & (0.0629) \\
Entrepreneurship & 0.1856** & 0.1839*** & -0.0748 & 0.0771 \\
 & (0.0799) & (0.0652) & (0.1009) & (0.0924) \\
Executive Compensation & 0.1670** & 0.1798** & -0.2195** & 0.0302 \\
 & (0.0783) & (0.0813) & (0.0970) & (0.0980) \\
Family Business & -0.1898*** & -0.1736** & 0.2753*** & 0.1513* \\
 & (0.0707) & (0.0682) & (0.0834) & (0.0805) \\
Firm Strategy & 0.1509* & -0.0228 & -0.3518*** & -0.8263*** \\
 & (0.0820) & (0.0722) & (0.0857) & (0.0753) \\
Game Theory & 0.0196 & 0.1663*** & -0.1058 & 0.1186 \\
 & (0.0620) & (0.0610) & (0.0751) & (0.0882) \\
Gender Economics & -0.7051*** & -0.7374*** & 0.7767*** & 0.9348*** \\
 & (0.0834) & (0.0678) & (0.1027) & (0.1116) \\
Health Economics & -0.3489*** & -0.2133*** & 0.5279*** & 0.4761*** \\
 & (0.0563) & (0.0436) & (0.0595) & (0.0591) \\
Hedge Funds & 0.1050 & -0.0190 & 0.0026 & 0.0517 \\
 & (0.0776) & (0.0690) & (0.0698) & (0.1030) \\
Innovation and Patents & 0.0799 & 0.2205*** & 0.1933*** & 0.3388*** \\
 & (0.0594) & (0.0585) & (0.0700) & (0.0580) \\
Intergenerational Mobility & -0.0745 & -0.2183*** & -0.1361* & -0.1150 \\
 & (0.0629) & (0.0544) & (0.0706) & (0.0749) \\
International Business & 0.0350 & 0.1392** & -0.2061*** & -0.0903 \\
 & (0.0497) & (0.0640) & (0.0605) & (0.0654) \\
Labor and Wages & 0.4273*** & 0.2527*** & -0.3033*** & -0.5526*** \\
 & (0.0964) & (0.0882) & (0.0920) & (0.1127) \\
Mergers and Acquisitions & 0.3363*** & 0.2849*** & -0.1417 & -0.3722*** \\
 & (0.0808) & (0.0816) & (0.0909) & (0.1015) \\
Online Reviews & 0.4353*** & 0.3191*** & -0.2358*** & -0.3654*** \\
 & (0.0594) & (0.0616) & (0.0439) & (0.0634) \\
Personnel Economics & 0.1700** & 0.2268** & -0.4815*** & -0.2085*** \\
 & (0.0702) & (0.0984) & (0.0571) & (0.0703) \\
Platform Economics & -0.1717** & -0.1267 & 0.4809*** & 0.3304*** \\
 & (0.0860) & (0.0795) & (0.0887) & (0.0813) \\
Public Governance & 0.0973** & 0.0682 & -0.1350** & -0.2045*** \\
 & (0.0416) & (0.0538) & (0.0655) & (0.0541) \\
Racial Inequality & -0.1463*** & 0.1735*** & 0.4108*** & 0.5100*** \\
 & (0.0466) & (0.0573) & (0.0581) & (0.0493) \\
Retail Pricing & 0.2590** & 0.2154* & -0.2617** & -0.5095*** \\
 & (0.1121) & (0.1163) & (0.1243) & (0.1279) \\
Risk Preferences & 0.2463*** & 0.1724* & -0.2831*** & -0.3456*** \\
 & (0.0771) & (0.0880) & (0.0694) & (0.0744) \\
Supply Chain Economics & -0.2434** & -0.2505** & 0.0793 & 0.4925*** \\
 & (0.1077) & (0.1036) & (0.1165) & (0.1141) \\
Team Productivity & 0.2257*** & 0.1526*** & -0.3122*** & -0.4037*** \\
 & (0.0585) & (0.0584) & (0.0713) & (0.0673) \\
Venture Capital & -0.1673 & -0.1829*** & -0.0747 & -0.1758 \\
 & (0.1097) & (0.0701) & (0.1226) & (0.1134) \\
\midrule
AI Topicality $\times$ Year Terms & Yes & Yes & Yes & Yes \\
Forum Indicators & Yes & Yes & Yes & Yes \\
Year Fixed Effects & Yes & Yes & Yes & Yes \\
Thread-Title Emotion Controls & Yes & Yes & Yes & Yes \\
Observations & 1,255,942 & 1,255,942 & 1,255,942 & 1,255,942 \\
Pseudo $R^2$ & 0.0022 & 0.0018 & 0.0044 & 0.0043 \\
\bottomrule
\end{longtable}

\noindent\footnotesize Notes: Dependent variables are reply-level sentiment scores on the $[0,1]$ scale. The first two columns measure openness and curiosity, for which higher values indicate greater openness and curiosity; the second two columns measure negative tone and poisonousness, for which higher values indicate more negative and poisonous language. Standard errors are two-way clustered by calendar year and thread. * $p<0.1$, ** $p<0.05$, *** $p<0.01$.
\normalsize

\subsection{Additional Descriptive Statistics}
\begin{center}
\scriptsize
\setlength{\tabcolsep}{4pt}
\begin{longtable}{lrrrrr}
\caption{Descriptive Statistics for Control Variables in the Appendix}\label{tab:descriptive_controls_appendix}\\
\toprule
Variable & N & Mean & Std. Dev. & Min & Max \\
\midrule
\endfirsthead
\multicolumn{6}{l}{\tablename\ \thetable{} -- continued from previous page}\\
\toprule
Variable & N & Mean & Std. Dev. & Min & Max \\
\midrule
\endhead
\midrule
\multicolumn{6}{r}{Continued on next page}\\
\endfoot
\bottomrule
\endlastfoot
\addlinespace
\multicolumn{6}{l}{\textit{Thread-Title Emotion Controls}} \\
Thread Admiration & 1,255,942 & 0.0351 & 0.1464 & 0.0004 & 0.9585 \\
Thread Amusement & 1,255,942 & 0.0083 & 0.0655 & 0.0002 & 0.9565 \\
Thread Anger & 1,255,942 & 0.0104 & 0.0534 & 0.0002 & 0.8517 \\
Thread Annoyance & 1,255,942 & 0.0304 & 0.0762 & 0.0011 & 0.8037 \\
Thread Approval & 1,255,942 & 0.0343 & 0.0707 & 0.0017 & 0.9252 \\
Thread Caring & 1,255,942 & 0.0048 & 0.0355 & 0.0002 & 0.9157 \\
Thread Confusion & 1,255,942 & 0.0951 & 0.1500 & 0.0003 & 0.9412 \\
Thread Curiosity & 1,255,942 & 0.1841 & 0.2656 & 0.0002 & 0.9069 \\
Thread Desire & 1,255,942 & 0.0075 & 0.0540 & 0.0003 & 0.8438 \\
Thread Disappointment & 1,255,942 & 0.0223 & 0.0766 & 0.0005 & 0.8670 \\
Thread Disapproval & 1,255,942 & 0.0256 & 0.0864 & 0.0004 & 0.8974 \\
Thread Disgust & 1,255,942 & 0.0055 & 0.0331 & 0.0002 & 0.8480 \\
Thread Embarrassment & 1,255,942 & 0.0023 & 0.0173 & 0.0001 & 0.8128 \\
Thread Excitement & 1,255,942 & 0.0075 & 0.0359 & 0.0002 & 0.8384 \\
Thread Fear & 1,255,942 & 0.0045 & 0.0393 & 0.0001 & 0.9115 \\
Thread Gratitude & 1,255,942 & 0.0025 & 0.0299 & 0.0001 & 0.9904 \\
Thread Grief & 1,255,942 & 0.0008 & 0.0027 & 0.0001 & 0.0837 \\
Thread Joy & 1,255,942 & 0.0078 & 0.0495 & 0.0003 & 0.9134 \\
Thread Love & 1,255,942 & 0.0061 & 0.0543 & 0.0001 & 0.9641 \\
Thread Nervousness & 1,255,942 & 0.0018 & 0.0136 & 0.0000 & 0.6433 \\
Thread Optimism & 1,255,942 & 0.0076 & 0.0327 & 0.0005 & 0.9336 \\
Thread Pride & 1,255,942 & 0.0007 & 0.0045 & 0.0000 & 0.4629 \\
Thread Realization & 1,255,942 & 0.0157 & 0.0441 & 0.0015 & 0.8787 \\
Thread Relief & 1,255,942 & 0.0010 & 0.0042 & 0.0001 & 0.2478 \\
Thread Remorse & 1,255,942 & 0.0019 & 0.0252 & 0.0001 & 0.8293 \\
Thread Sadness & 1,255,942 & 0.0156 & 0.0773 & 0.0003 & 0.9264 \\
Thread Surprise & 1,255,942 & 0.0071 & 0.0426 & 0.0002 & 0.8987 \\
Thread Neutral & 1,255,942 & 0.5945 & 0.3242 & 0.0045 & 0.9737 \\
\addlinespace
\multicolumn{6}{l}{\textit{Forum Indicators}} \\
Forum China-Job-Market & 1,255,942 & 0.0064 & 0.0798 & 0.0000 & 1.0000 \\
Forum Conferences & 1,255,942 & 0.0013 & 0.0356 & 0.0000 & 1.0000 \\
Forum Econ-Lounge & 1,255,942 & 0.0249 & 0.1557 & 0.0000 & 1.0000 \\
Forum Econometrics-Discussion & 1,255,942 & 0.0386 & 0.1927 & 0.0000 & 1.0000 \\
Forum Economics-Discussion & 1,255,942 & 0.1587 & 0.3654 & 0.0000 & 1.0000 \\
Forum European-Job-Market & 1,255,942 & 0.0024 & 0.0494 & 0.0000 & 1.0000 \\
Forum Finance-Job-Rumors & 1,255,942 & 0.0477 & 0.2132 & 0.0000 & 1.0000 \\
Forum From-The-Blogs & 1,255,942 & 0.0057 & 0.0754 & 0.0000 & 1.0000 \\
Forum General-Economics-Job-Market-Discussion & 1,255,942 & 0.0388 & 0.1930 & 0.0000 & 1.0000 \\
Forum Industry-Rumors & 1,255,942 & 0.0062 & 0.0783 & 0.0000 & 1.0000 \\
Forum Latest-Research-Discussion & 1,255,942 & 0.0101 & 0.1000 & 0.0000 & 1.0000 \\
Forum Macro-Job-Rumors & 1,255,942 & 0.0014 & 0.0377 & 0.0000 & 1.0000 \\
Forum Macroeconomics & 1,255,942 & 0.0081 & 0.0898 & 0.0000 & 1.0000 \\
Forum Math-Job-Market & 1,255,942 & 0.0042 & 0.0644 & 0.0000 & 1.0000 \\
Forum Math-Lounge-Off-Topic & 1,255,942 & 0.0184 & 0.1344 & 0.0000 & 1.0000 \\
Forum Micro-Job-Rumors & 1,255,942 & 0.0015 & 0.0390 & 0.0000 & 1.0000 \\
Forum Microeconomics & 1,255,942 & 0.0040 & 0.0634 & 0.0000 & 1.0000 \\
Forum Off-Topic & 1,255,942 & 0.4985 & 0.5000 & 0.0000 & 1.0000 \\
Forum Political-Economy & 1,255,942 & 0.0153 & 0.1228 & 0.0000 & 1.0000 \\
Forum Political-Science-Discussion & 1,255,942 & 0.0002 & 0.0147 & 0.0000 & 1.0000 \\
Forum Political-Science-Job-Market & 1,255,942 & 0.0000 & 0.0039 & 0.0000 & 1.0000 \\
Forum Political-Science-Lounge-Off-Topic & 1,255,942 & 0.0001 & 0.0108 & 0.0000 & 1.0000 \\
Forum Questions-From-Prospective-Grad-Students & 1,255,942 & 0.0181 & 0.1332 & 0.0000 & 1.0000 \\
Forum Registered-Users-Forum & 1,255,942 & 0.0000 & 0.0013 & 0.0000 & 1.0000 \\
Forum Research-Journals & 1,255,942 & 0.0217 & 0.1459 & 0.0000 & 1.0000 \\
Forum Sociology-Discussion & 1,255,942 & 0.0017 & 0.0411 & 0.0000 & 1.0000 \\
Forum Sociology-Job-Market & 1,255,942 & 0.0000 & 0.0061 & 0.0000 & 1.0000 \\
Forum Sociology-Lounge-Off-Topic & 1,255,942 & 0.0000 & 0.0062 & 0.0000 & 1.0000 \\
Forum Software-And-Programming-For-Research & 1,255,942 & 0.0118 & 0.1080 & 0.0000 & 1.0000 \\
Forum Sport & 1,255,942 & 0.0031 & 0.0559 & 0.0000 & 1.0000 \\
Forum Teaching & 1,255,942 & 0.0102 & 0.1004 & 0.0000 & 1.0000 \\
Forum Technology & 1,255,942 & 0.0397 & 0.1952 & 0.0000 & 1.0000 \\
Forum Trash & 1,255,942 & 0.0010 & 0.0310 & 0.0000 & 1.0000 \\
\end{longtable}
\noindent\footnotesize Notes: The table reports summary statistics for the thread-title emotion controls and forum indicators used in the regression specifications. Emotion controls are scores on the $[0,1]$ scale; forum indicators are binary variables.
\normalsize
\end{center}

\end{document}